\documentclass[prd,aps,nofootinbib,notitlepage,showpacs,preprintnumbers]{revtex4-1}
\usepackage{graphicx,epsf,amsmat
h,amsfonts,amssymb,amsbsy}
\usepackage{epsfig}
\newcommand{\ds}{\displaystyle}
\newcommand{\vev}[1]{\langle#1\rangle}
\newcommand{\mat}{\left ( \begin{array}}
\newcommand{\emat}{\end{array} \right )}
\newcommand{\vect}{\left ( \begin{array}{c}}
\newcommand{\evect}{\end{array} \right )}

\begin{document}


\title{Competition and duality correspondence between chiral and
  superconducting channels in (2+1)-dimensional four-fermion models
  with fermion number and chiral chemical potentials}
\author{D. Ebert $^{1)}$,
T.G. Khunjua $^{2)}$, K.G. Klimenko $^{3)}$, and V.C. Zhukovsky $^{2)}$}
\vspace{1cm}

 \affiliation{$^{1)}$ Institute of Physics,
Humboldt-University Berlin, 12489 Berlin, Germany}
\affiliation{$^{2)}$ Faculty of Physics, Moscow State University,
119991, Moscow, Russia} \affiliation{$^{3)}$ State Research Center
of Russian Federation -- Institute for High Energy Physics,
NRC "Kurchatov Institute", 142281, Protvino, Moscow Region, Russia}

\begin{abstract}
In this paper the duality correspondence between fermion-antifermion
and difermion interaction channels is established in two
(2+1)-dimensional Gross-Neveu type models with a fermion number
chemical potential $\mu$ and a chiral chemical potential $\mu_5$. The role and influence of this property on the phase structure of the models are investigated. In particular, it is shown that the
chemical potential $\mu_5$ promotes the appearance of dynamical
chiral symmetry breaking, whereas the chemical potential $\mu$
contributes to the emergence of superconductivity.
\end{abstract}


\maketitle

\section{Introduction}

It is well known that relativistic quantum field theory models with
four-fermion (4F) interactions serve as effective theories for low
energy considerations of different real phenomena in a variety of
physical branches. For example, the meson spectroscopy, neutron star
and heavy-ion collision physics are often investigated in the
framework of (3+1)-dimensional 4F theories \cite{volkov,buballa},
known as Nambu--Jona-Lasinio (NJL) models \cite{njl}. In particular,
low dimensional 4F field theories provide a powerful tool for
investigations in condensed-matter physics. Indeed, physics of
(quasi)one-dimensional organic Peierls insulators (the best known
material of this kind is polyacetylene) is well described in terms
of the (1+1)-dimensional 4F Gross-Neveu (GN) model
\cite{gn,Krive:1987cr,n2a,Chodos:1997pg,n2c}. The quasirelativistic
treatment of electrons in planar systems like high-temperature
superconductors or in graphene (a planar monoatomic layer of carbon
atoms) is also possible in terms of (2+1)-dimensional GN models
\cite{Semenoff:1998bk,zkke,marino,kzz,kzz2,n21,Ebert:2015hva}.

Notice that there are physical effects, which were observed for the
first time just in the framework of NJL- and GN-type models. In
particular, the phenomenon of a dynamical generation of a fermion
mass is well-known for strong interaction physics since the time,
when Nambu and Jona-Lasinio \cite{njl} proved a dynamical breaking
of the continuous $\gamma_5$-symmetry on the basis of a generic 4F
interaction theory. Later on, this effect served as the basis of a
qualitatively successful description of the low-energy meson
spectrum of quantum chromodynamics (QCD) \cite{volkov}. The effect
of dynamical symmetry breaking and generation of a fermion mass is
also known in low dimensional, D=1+1 and D=2+1, GN theories
\cite{gn,n11,24,28}, where the four-fermion theory is renormalizable
and asymptotic free \cite{gn} in the case of D=1+1, whereas for
D=2+1 the 4F GN models are perturbatively nonrenormalizable but
become renormalizable in the framework of the $1/N$ expansion
technique \cite{28} ($N$ being the number of fermion fields).
Another example of this kind is the  effect of spontaneous chiral
symmetry breaking induced by external magnetic or chromomagnetic
fields. This effect was for the first time studied also in terms of
the (2+1)-dimensional GN model \cite{klimenko}.

Note in addition that due to a rather simple structure, different
low dimensional GN models provide a good laboratory for a deeper
study of dense baryon matter and, in particular, for the
consideration of such phenomena as (color)superconductivity
\cite{chodos,toki}, charged pion condensation \cite{gubina1,gubina},
etc. Moreover, these theories are very useful in developing new
quantum field theoretical techniques like the $1/N$ expansion
\cite{28}, the optimized perturbation theory \cite{kneur,k}, and so
on.

It is necessary to point out that there is one more, not yet
mentioned above, phenomenon that is still a well-known feature of
only some (1+1)-dimensional 4F theories. This is the duality
correspondence between chiral symmetry breaking and
superconductivity \cite{oj,vas,thies1,ekkz}. (In order to avoid the
prohibition of Cooper pairing as well as spontaneous breaking of a
continuous symmetry in (1+1)-dimensional models \cite{coleman}, the
consideration should be performed in the leading order of the
$1/N$-technique, i.e. in the large-$N$ limit assumption. In this
case quantum fluctuations, which would otherwise destroy a
long-range order corresponding to spontaneous symmetry breaking, are
suppressed by $1/N$ factors.) To formulate this phenomenon, let us
imagine that there is a microscopic 4F theory (for D=1+1 see, e.g.,
\cite{chodos,ekkz}) which describes a competition between
fermion-antifermion (or chiral) and difermion (or superconducting)
channels of interaction. Moreover, we suppose that the ground state
of the model is characterized by nonzero fermion number- and chiral
charge densities (the last is an imbalance between densities of
left- and right-handed fermions), i.e. there are two external
parameters, fermion number- $\mu$ and chiral charge- $\mu_5$
chemical potentials, respectively. Then the duality phenomenon
means that the phase structure of this 4F model is symmetric with
respect to the following simultaneous transformations
$G_1\leftrightarrow G_2$, $\mu\leftrightarrow\mu_5$ and
chiral symmetry breaking $\leftrightarrow$ superconductivity (here
$G_1$ and $G_2$ are the coupling constants in the chiral and
superconducting channels, respectively). Thus, there is a
correspondence between properties (phase structure) of the model,
e.g., at $G_1<G_2$ and $G_1>G_2$, etc. Moreover, knowing condensates
and other dynamical and thermodynamical quantities of the system,
e.g. in the chirally broken phase, one can obtain the corresponding
quantities in the superconducting phase of the model, by simply
performing there the duality transformation.

It is worth to note that in recent years properties of media with
nonzero chiral chemical potential $\mu_5$, i.e. chiral media,
attracted considerable interest (see, e.g., \cite{andrianov,Braguta}
and references therein). In nature, chiral media might be realized
in heavy-ion collisions, compact stars, condensed matter systems,
etc. \cite{andrianov} (see also the review \cite{ms}).
In particular, one can expect that in the quark-gluon-plasma phase of QCD a chirality imbalance is produced, i.e. there appears a nonzero chiral charge density $n_5$. The combined effect of the chiral imbalance and of the very strong magnetic field, which can be produced at noncentral heavy ion collision, results in the so-called chiral magnetic effect (see, e.g., the paper \cite{kharzeev}). It means that there might be induced an electric current along the direction of the magnetic field. It is important to note that this phenomenon is described effectively in the framework of 4F models with a chiral chemical potential $\mu_5$, i.e. in terms of a quantity conjugated to the chiral charge density $n_5$.
Recently, it was established that in planar condensed matter systems there may exists a phenomenon, which can be regarded as an analogue of the QCD chiral magnetic effect \cite{mizher}. This is the pseudo chiral magnetic effect, which can be observed, e.g., in a distorted graphene sheet under the influence of external in-plane magnetic field. Indeed, due to a mechanical distortion of the lattice structure of graphene, there might appears a nonzero chiral charge density $n_5$ (chiral imbalance) as well as a conserved electric current along the in-plane external magnetic field (for details, see the paper \cite{mizher}). Notice that, similar to the QCD chiral magnetic effect, one might expect that the pseudo chiral magnetic effect can  be effectively described in terms of (2+1)-dimensional 4F models with nonzero chiral chemical potential $\mu_5$ (in addiion to the usual chemical potential $\mu$). Finally, it is important to remark that a mechanical distortion of the lattice structure in graphene-like materials (as well as other external perturbations) can also lead to opening of different superconducting channels in the system \cite{mudry}.

Thus, it makes sense to study a competition between chiral symmetry breaking and superconductivity in (2+1)-dimensional quantum field theories with chiral chemical potential $\mu_5$.  In particular, it is important to
predict/observe the possible duality relations between these qualitatively
different phenomena accompanied by $\mu_5$ in field theories in
spacetime dimensions D$>$1+1.

In this paper we demonstrate that there exists a dual correspondence
(or dual symmetry) between the phenomena of the chiral symmetry breaking and superconductivity in the framework of some (2+1)-dimensional 4F
models. The consideration is performed at zero temperature $T$, however it can be easily generalized to the case $T\ne 0$. We hope
that our investigations shed some new light on the physical effects
in planar condensed matter systems.

The paper is organized as follows. In Sec. II a (2+1)-dimensional 4F
GN-type model, containing both fermion-antifermion (or chiral) and
difermion (or superconducting) interaction channels and including
two kinds of chemical potentials, $\mu,\mu_5$, is presented. Here we
show that under the Pauli-G\"ursey transformations of fermi fields
there is a dual relationship between the relevant 4F structures.
Next, the unrenormalized thermodynamic potential (TDP) of the
GN-type model is given in the leading order of the large-N
expansion. In Sec. III the dual symmetry of the model TDP is
established. It means that it is invariant under the interchange of
coupling constants, $\mu,\mu_5$ chemical potentials and chiral and
superconducting condensates. Moreover, the renormalization of the
TDP is performed. Sec. IV contains a detailed numerical investigation
of various  phase portraits with particular emphasis on the role of
the duality symmetry of the TDP. Some technical details are
relegated to Appendices. Moreover, in Appendix \ref{ApD} we present
an alternative (2+1)-dimensional GN model with dual relationship
between other chiral and superconducting channels of 4F interaction.

\section{ The model and its thermodynamic potential}
\label{effaction}

Our investigation is based on a (2+1)-dimensional GN--type model
with massless fermions belonging to a fundamental multiplet of the
auxiliary $O(N)$ flavor group. Its Lagrangian describes the
interaction both in the fermion--antifermion and difermion (or
superconducting) channels:
\begin{eqnarray}
 L\equiv L(G_1,G_2;\mu,\mu_5)=\sum_{k=1}^{N}\bar \psi_k\Big [\gamma^\nu i\partial_\nu
+\mu\gamma^0+\mu_5\gamma^0\gamma^5\Big ]\psi_k&+& \frac {G_1}N\left (4F\right )_{ch}+\frac {G_2}N\left
(4F\right )_{sc},\label{1}
\end{eqnarray}
where the four-fermion structures $\left (4F\right )_{ch}$ and $(4F)_{sc}$ are used,
\begin{eqnarray}
\left (4F\right )_{ch}=\left
(\sum_{k=1}^{N}\bar \psi_k\psi_k\right )^2+\left (\sum_{k=1}^{N}\bar \psi_k
i\gamma^5\psi_k\right )^2,~~(4F)_{sc}=\left
(\sum_{k=1}^{N} \psi_k^T C\psi_k\right )\left (\sum_{j=1}^{N}\bar
\psi_j C\bar\psi_j^T\right ). \label{01}
\end{eqnarray}
In addition, $\mu$ and $\mu_5$ in (1) denote a fermion number
chemical potential and a chiral (axial) chemical potential, respectively.
$\mu$ is conjugated to a fermion number density
$n$, whereas $\mu_5$ is conjugated to a nonzero density of chiral
charge $n_5=n_R-n_L$, which represents an imbalance in densities of
the right- and left-handed fermions. As it is noted above, all
fermion fields $\psi_k$ ($k=1,...,N$) form a fundamental multiplet
of the $O(N)$ group. Moreover, each field $\psi_k$ is a
four-component (reducible) Dirac spinor (the symbol $T$ denotes the
transposition operation). Therefore, in Eqs. (1)-(2) the quantities
$\gamma^\nu$ ($\nu =0,1,2$) and $\gamma^5$ are matrices in the
four-dimensional spinor space. Moreover, $C\equiv\gamma^2$ is the
charge conjugation matrix. The algebra of these matrices as well as
their particular representations are given in Appendix \ref{ApA}.
Clearly, the Lagrangian $L$ is invariant under transformations from
the internal auxiliary $O(N)$ group, which is introduced here in
order to make it possible to perform all the calculations in the
framework of the nonperturbative large-$N$ expansion method.
Physically more interesting is that the model (1) is invariant under
transformations from the $U_V(1)\times U_{\gamma^5}(1)$ group, where
$U_V(1)$ is the fermion number conservation group, $\psi_k\to\exp
(i\alpha)\psi_k$ ($k=1,..,N$), and $U_{\gamma^5}(1)$ is the group of
continuous chiral transformations, $\psi_k\to\exp
(i\alpha\gamma^5)\psi_k$ ($k=1,...,N$). It means that in the
framework of the model (1) both the particle number density
$n=\sum_{k=1}^{N}\bar \psi_k\gamma^0\psi_k$ and the density of
chiral charge $n_5=\sum_{k=1}^{N}\bar \psi_k\gamma^0\gamma^5\psi_k$
are conserved quantities. \footnote{Since for the reducible four-component
spinor representation there is one more Hermitian
matrix $\gamma^3$, which anticommutes with $\gamma^\nu$ ($\nu
=0,1,2$) and $\gamma^5$, one can consider another continuous chiral
$U_{\gamma^3}(1)$ transformation group of the spinor fields,
$\psi_k\to\exp(i\alpha\gamma^3)\psi_k$ ($k=1,...,N$). Alternatively,
there exists a 4F model with another fermion--antifermion and
difermion channels of interaction, which, in addition to $U_V(1)$,
is invariant under the continuous chiral $U_{\gamma^3}(1)$ group
(see Appendix \ref{ApD}). It is worth mentioning that in the case of
graphene reducible four-component spinors just describe the two
sublattice (or pseudospin) and two valley (or Dirac point) degrees
of freedom of the hexagonal honeycomb lattice of carbon atoms (see
e.g. \cite{Ebert:2015hva} and references therein).}

Before studying the thermodynamics of the model, we would like to
point out that there is a so-called duality correspondence between
chiral and superconducting channels of the model (1)-(2).  To see
this, it is very useful to form an infinite set $\boldsymbol {\cal
F}$ composed of all Lagrangians $ L(G_1,G_2;\mu,\mu_5)$ when the
free model parameters $G_1,G_2,\mu$ and $\mu_5$  take arbitrary
admissible values, i.e. ${\cal L}(G_1,G_2;\mu,\mu_5)\in \boldsymbol
{\cal F}$ at arbitrary fixed values of coupling constants $G_1>0$,
 $G_2>0$ and chemical potentials $\mu, \mu_5$. Then, let us perform in
(1)-(2) the so-called Pauli-G\"ursey (PG) transformation of spinor
fields \cite{pauli},
 \begin{eqnarray}
PG:~~\psi_k (x)
  \longrightarrow \frac 12 (1-\gamma^5)\psi_k (x)+\frac 12
(1+\gamma^5)C\bar\psi^T_k(x). \label{0004}
\end{eqnarray}
Taking into account that all spinor fields anticommute with each
other, it is easy to see that under the action of the
transformations (\ref{0004}) the 4F structures of the Lagrangian (1)
are converted into themselves, i.e.
\begin{eqnarray}
(4F)_{ch}\stackrel{
  PG}{\longleftrightarrow} (4F)_{sc}, \label{040}
\end{eqnarray}
and, moreover, each element (Lagrangian) $ L(G_1,G_2;\mu,\mu_5)$ of the set $\boldsymbol {\cal F}$ is transformed into another element of the set
$\boldsymbol {\cal F}$ according to the following rule
\begin{eqnarray}
L(G_1,G_2;\mu,\mu_5) \stackrel{ PG}{\longleftrightarrow}
L(G_2,G_1;-\mu_5,-\mu)\in \boldsymbol {\cal F}, \label{004}
\end{eqnarray}
i.e. the set $\boldsymbol {\cal F}$ is invariant under the field
transformations (\ref{0004}). Owing to the relations  (\ref{040})
and (\ref{004}) there is a connection between properties of the
model at some fixed free model parameters $G_1,G_2,\mu,\mu_5$  and
properties of the model in the case, when $G_1\leftrightarrow G_2$
and $\mu\leftrightarrow\mu_5$. In particular, if at some fixed
$G_1,G_2,\mu$ and $\mu_5$ we have the chiral symmetry breaking (CSB)
phase, then at $G_1\leftrightarrow G_2$ and
$\mu\leftrightarrow\mu_5$ one can definitely predict the
superconducting (SC) phase, and vice versa. Due to this reason, we
will call the relations (\ref{040}) and (\ref{004}) the duality
property of the model (or duality correspondence between CSB and
SC). Note also that in Appendix \ref{ApD} we present another
(2+1)-dimensional 4F model, in which the duality correspondence
between alternative chiral and superconducting channels is realized.

Further on, we will study the role and the influence of the duality
property of the model (1) on its phase structure.  To this end, we
introduce the semi-bosonized version of Lagrangian (\ref{1}) that
contains only quadratic powers of fermionic fields as well as
auxiliary bosonic fields $\sigma (x)$, $\pi(x)$, $\Delta(x)$ and
$\Delta^{*}(x)$,
\begin{eqnarray}
{\cal L}\ds =\bar\psi_k\Big [\gamma^\nu i\partial_\nu +\mu\gamma^0
+\mu_5\gamma^0\gamma^5-\sigma -i\gamma^5\pi\Big ]\psi_k
 -\frac{N(\sigma^2+\pi^2)}{4G_1} -\frac N{4G_2}\Delta^{*}\Delta-
 \frac{\Delta^{*}}{2}[\psi_k^TC\psi_k]
-\frac{\Delta}{2}[\bar\psi_k C\bar\psi_k^T]. \label{2}
\end{eqnarray}
(Here and in what follows the summations over repeated indices
$k=1,...,N$ and $\nu=0,1,2$ are implied.) Clearly, the Lagrangians
(1) and (\ref{2}) are equivalent, as can be seen by using the
Euler-Lagrange equations of motion for  bosonic fields which take
the form
\begin{eqnarray}
\sigma (x)=-2\frac {G_1}N(\bar\psi_k\psi_k),~~\pi(x)=-2\frac {G_1}N(\bar\psi_k i\gamma^5\psi_k),~~ \Delta(x)=-2\frac
{G_2}N(\psi_k^TC\psi_k),~~ \Delta^{*}(x)=-2\frac
{G_2}N(\bar\psi_k C\bar\psi_k^T). \label{3}
\end{eqnarray}
One can easily see from (\ref{3}) that the neutral fields
$\sigma(x)$ and $\pi(x)$ are real quantities, i.e.
$(\sigma(x))^\dagger=\sigma(x)$ and $(\pi(x))^\dagger=\pi(x)$ (the
superscript symbol $\dagger$ denotes the Hermitian conjugation), but
the (charged) difermion fields $\Delta(x)$ and $\Delta^*(x)$ are
Hermitian conjugated complex quantities, so that
$(\Delta(x))^\dagger= \Delta^{*}(x)$ and vice versa. Moreover, under
the chiral $U_{\gamma^5}(1)$ group the fields
$\Delta(x),\Delta^{*}(x)$ are singlets, but the fields $\sigma
(x),\pi(x)$ are transformed in the following way
\begin{eqnarray}
U_{\gamma^5}(1):~
\sigma (x)\to\cos(2\alpha)\sigma (x)+\sin(2\alpha)\pi(x),~
\pi (x)\to-\sin(2\alpha)\sigma (x)+\cos(2\alpha)\pi(x).\label{003}
\end{eqnarray}
Clearly, all the fields (\ref{3}) are also singlets with respect to the
auxiliary $O(N)$ group, since the representations of this group are real.
Moreover, with respect to the parity transformation $P$,
\begin{eqnarray}
P:~~\psi_k (t,x,y)\to i\gamma^5\gamma^1\psi_k (t,-x,y),~~~~k=1,...,N,
\label{03}
\end{eqnarray}
the fields $\sigma(x)$, $\Delta (x)$ and $\Delta^{*}(x)$ are even quantities, i.e. scalars, but $\pi(x)$ is a pseudoscalar. If the
difermion field $\Delta(x)$ has a nonzero ground state expectation value,
i.e.\  $\vev{\Delta(x)}\ne 0$, then the Abelian fermion number conservation $U_V(1)$ symmetry of the model is spontaneously broken down and the superconducting phase is realized in the model. However, if
$\vev{\sigma (x)}\ne 0$ then the continuous $U_{\gamma^5}(1)$ chiral
symmetry of the model is spontaneously broken.

Let us now study the phase structure of the four-fermion model (1)
starting from the equivalent semi-bosonized Lagrangian (\ref{2}).
In the leading order of the large-$N$ approximation, the effective
action ${\cal S}_{\rm {eff}}(\sigma,\pi,\Delta,\Delta^{*})$ of the
considered model is expressed by means of the path integral over
fermion fields:
$$
\exp(i {\cal S}_{\rm {eff}}(\sigma,\pi,\Delta,\Delta^{*}))=
  \int\prod_{l=1}^{N}[d\bar\psi_l][d\psi_l]\exp\Bigl(i\int {\cal
  L}\,d^3 x\Bigr),
$$
leading to
\begin{eqnarray}
&&{\cal S}_{\rm {eff}} (\sigma,\pi,\Delta,\Delta^{*}) =-\int
d^3x\left [\frac{N(\sigma^2(x)+\pi^2(x))}{4G_1}+
\frac{N}{4G_2}\Delta (x)\Delta^{*}(x)\right ]+ \widetilde {\cal
S}_{\rm {eff}}.  \label{5}
\end{eqnarray}
The term $\widetilde {\cal S}_{\rm {eff}}$ in (\ref{5})
is the fermion contribution to the effective action
and is given by:
\begin{equation}
\exp(i\widetilde {\cal S}_{\rm
{eff}})=\int\prod_{l=1}^{N}[d\bar\psi_l][d\psi_l]\exp\Bigl\{i\int\Big
[\bar\psi_k(\gamma^\nu i\partial_\nu +\mu\gamma^0+\mu_5\gamma^0\gamma^5 -\sigma-i\gamma^5\pi )\psi_k -
 \frac{\Delta^{*}}{2}(\psi_k^T C\psi_k)
-\frac{\Delta}{2}(\bar\psi_kC \bar\psi_k^T)\Big ]d^3
x\Bigr\}. \label{6}
\end{equation}
The ground state expectation values $\vev{\sigma(x)}$,
$\vev{\Delta(x)}$, etc. of the composite bosonic fields are determined
by the saddle point equations,
\begin{eqnarray}
\frac{\delta {\cal S}_{\rm {eff}}}{\delta\sigma (x)}=0,~~~~~
\frac{\delta {\cal S}_{\rm {eff}}}{\delta\pi (x)}=0,~~~~~
\frac{\delta {\cal S}_{\rm {eff}}}{\delta\Delta (x)}=0,~~~~~
\frac{\delta {\cal S}_{\rm {eff}}}{\delta\Delta^* (x)}=0. \label{7}
\end{eqnarray}
For simplicity, throughout the paper we suppose that the above
mentioned ground state expectation values do not depend on spacetime
coordinates, i.e.
\begin{eqnarray}
\vev{\sigma(x)}\equiv M,~~~\vev{\pi(x)}\equiv \pi,\vev{\Delta(x)}\equiv \Delta,~~~
\vev{\Delta^*(x)}\equiv \Delta^*, \label{8}
\end{eqnarray}
where $M,\pi,\Delta,\Delta^*$ are constant quantities. In fact,
they are coordinates of the global minimum point of the
thermodynamic potential (TDP) $\Omega (M,\pi,\Delta,\Delta^*)$.
In the leading order of the large-$N$ expansion and using (13)
it is defined by the following expression:
\begin{equation}
\int d^3x \Omega (M,\pi,\Delta,\Delta^*)=-\frac{1}{N}{\cal S}_{\rm
{eff}}\big (\sigma(x),\pi (x),\Delta (x),\Delta^*(x)\big )\Big|_{\sigma
(x)=\vev{\sigma(x)},\Delta(x)=\vev{\Delta(x)},...} .\label{08}
\end{equation}
The TDP (\ref{08}) is invariant with respect to chiral $U_{\gamma^5}(1)$
symmetry group. So, as it is clear from (\ref{003}), it depends on the
quantities $M$ and $\pi$ through the combination $M^2+\pi^2$. Moreover,
without loss of generality, one can suppose that $\vev{\pi(x)}\equiv
\pi=0$. Thus, to find the other ground state expectation values
$\vev{\sigma(x)}$ etc., it is enough to study the global minimum point of
the TDP $\Omega (M,\Delta,\Delta^*)$,
\begin{equation}
\Omega (M,\Delta,\Delta^*)\equiv\Omega (M,\pi,\Delta,\Delta^*)\Big|_{\pi=0} .\label{09}
\end{equation}
Taking into account the relations (\ref{5}), (\ref{6}) and (\ref{08}),
we have from (\ref{09})
\begin{eqnarray}
\int d^3x\Omega (M,\Delta,\Delta^*)\,\,&=&\,\,\int d^3x\left
(\frac{M^2}{4G_1}+\frac{\Delta\Delta^*}{4G_2}\right )+\frac{i}{N}\ln\left
( \int\prod_{l=1}^{N}[d\bar\psi_l][d\psi_l]\exp\Big (i\int d^3 x\Big
[\bar\psi_k D\psi_k\right.\nonumber\\&& \left.-
\frac{\Delta^*}{2}(\psi_k^TC\psi_k)
-\frac{\Delta}{2}(\bar\psi_k C\bar\psi_k^T)\Big ]
\Big )\right ), \label{9}
\end{eqnarray}
where $D=\gamma^\rho
i\partial_\rho+\mu\gamma^0+\mu_5\gamma^0\gamma^5-M$. To proceed
further, let us point out again that without loss of generality the
quantities $\Delta,\Delta^*$ might be considered as real
ones. \footnote{Otherwise,  phases of the complex values
  $\Delta,\Delta^*$ might be eliminated by an appropriate
  transformation of fermion fields in the path integral (\ref{9}).}
So, in what follows we will suppose that
$\Delta=\Delta^*\equiv\Delta$, where $\Delta$ now is already a real
quantity.

\section{Calculation of the TDP}

The path integration in the expression (\ref{9}) is evaluated in
Appendix \ref{ApB} \footnote{In Appendix \ref{ApB} we consider for
simplicity the case $N=1$, however the procedure is easily
generalized to the case with $N>1$.}, so we have for the TDP
(\ref{9}) the following expression
\begin{eqnarray}
\Omega (M,\Delta)\equiv\Omega^{un} (M,\Delta)&=&
\frac{M^2}{4G_1}+\frac{\Delta^2}{4G_2}
+\frac{i}{2}\int\frac{d^3p}{(2\pi)^3}\ln\Big
[\lambda_1(p)\lambda_2(p)\lambda_3(p)\lambda_4(p)\Big ], \label{11}
\end{eqnarray}
where $\lambda_{1,...,4}(p)$ are presented in (\ref{B12}) and
superscription ''un'' denotes the unrenormalized quantity. Note that the
TDP (\ref{11}) describes thermodynamics of the model at zero
temperature. Taking into account the expressions (\ref{B12}),
the TDP (\ref{11}) can be presented in the form
\begin{eqnarray}
\Omega^{un} (M,\Delta)&=&
\frac{M^2}{4G_1}+\frac{\Delta^2}{4G_2}
+\frac{i}{2}\sum_{\eta =\pm}\int\frac{d^3p}{(2\pi)^3}\ln P_\eta(p_0), \label{012}
\end{eqnarray}
where $P_\eta(p_0)\equiv a+\eta bp_0-2c p_0^2+p_0^4$ and
\begin{eqnarray}
a&=&(\mu_5^2-\mu^2+M^2-\Delta^2)^2-2|\vec p|^2(\mu_5^2+\mu^2-M^2-\Delta^2)+|\vec p|^4,\nonumber\\
b&=&8\mu\mu_5|\vec p|,~~c=\mu_5^2+|\vec p|^2+\mu^2+M^2+\Delta^2,~~|\vec p|=\sqrt{p_1^2+p_2^2}. \label{013}
\end{eqnarray}
It is clear from (\ref{013}) that the TDP (\ref{012}) is an even
function of each of the quantities $\mu$, $\mu_5$, $M$, and $\Delta$, i.e. without loss of generality we can consider in the following only $\mu\ge 0$, $\mu_5\ge 0$, $M\ge 0$, and $\Delta\ge 0$ values of these quantities. Moreover, the TDP (\ref{012}) is invariant with respect to the so-called duality transformation \cite{thies1,ekkz},
\begin{eqnarray}
{\cal D}:~~~~G_1\longleftrightarrow G_2,~~M\longleftrightarrow \Delta,~~\mu\longleftrightarrow\mu_5.
 \label{16}
\end{eqnarray}
Notice that this invariance of the TDP is a consequence of the rule
(\ref{004}), according to which the 4F Lagrangian (1) is modified
under the Pauli-Gursey transformation (\ref{0004}) of the spinor field.

In powers of $\Delta$ the fourth-degree polynomial $P_\eta(p_0)$ has the following form
\begin{eqnarray}
P_\eta(p_0)&\equiv&\Delta^4-2\Delta^2(p_0^2-|\vec p|^2+M^2+\mu_5^2-\mu^2)\nonumber\\
&+&\big (M^2+(|\vec p|-\mu_5)^2-(p_0-\eta\mu)^2\big )\big (M^2+(|\vec p|+\mu_5)^2-
(p_0+\eta\mu)^2\big ). \label{17}
\end{eqnarray}
Expanding the right-hand side of (\ref{17}) in powers of $M$, one can
obtain an equivalent alternative expression for this polynomial. Namely,
\begin{eqnarray}
P_\eta(p_0)&\equiv& M^4-2M^2(p_0^2-|\vec p|^2+\Delta^2+\mu^2-\mu_5^2)\nonumber\\
&+&\big (\Delta^2+(|\vec p|-\mu)^2-(p_0-\eta\mu_5)^2\big )\big
(\Delta^2+(|\vec p|+\mu)^2-(p_0+\eta\mu_5)^2\big ).\label{18}
\end{eqnarray}
Note also that according to the general theorem of algebra, the polynomial
$P_\eta(p_0)$ can be presented in the form
\begin{eqnarray}
P_\eta(p_0)\equiv (p_0-p^\eta_{01})(p_0-p^\eta_{02})(p_0-p^\eta_{03})(p_0-p^\eta_{04}), \label{170}
\end{eqnarray}
where $p^\eta_{01}$, $p^\eta_{02}$, $p^\eta_{03}$ and $p^\eta_{04}$
are the roots of this polynomial. The fourth-order polynomial
with similar coefficients $a,b,c$ (\ref{013}) was studied in our paper \cite{ekkz}, where it was shown that all its roots $p^\eta_{0i}$ ($i=1,...,4$) are real quantities (see Appendix B in \cite{ekkz}). The roots $p^\eta_{0i}$ are the energies of quasiparticle or quasiantiparticle excitations of the system. In particular, it follows from (\ref{17}) that at $\Delta=0$
\begin{eqnarray}
\Big (p^\eta_{01},p^\eta_{02}\Big )\Big |_{\Delta=0}=\eta\mu\pm\sqrt{M^2+(\mu_5-|\vec p|)^2},~~\Big (p^\eta_{03},p^\eta_{04}\Big )\Big |_{\Delta=0}=-\eta\mu\pm\sqrt{M^2+(\mu_5+|\vec p|)^2}, \label{26}
\end{eqnarray}
whereas it is clear from (\ref{18}) that at $M=0$
\begin{eqnarray}
\Big (p^\eta_{01},p^\eta_{02}\Big )\Big |_{M=0}=\eta\mu_5\pm\sqrt{\Delta^2+(\mu-|\vec p|)^2},~~\Big (p^\eta_{03},p^\eta_{04}\Big )\Big |_{M=0}=-\eta\mu_5\pm\sqrt{\Delta^2+(\mu+|\vec p|)^2}. \label{27}
\end{eqnarray}
Taking into account in (\ref{012}) the relation (\ref{170}) as well as the
formula
\begin{eqnarray}
\int_{-\infty}^\infty dp_0\ln\big
(p_0-K)=\mathrm{i}\pi|K|,\label{int}
\end{eqnarray}
(obtained rigorously, e.g., in Appendix B of \cite{gubina} and being
true up to an infinite term independent of the real quantity $K$),
it is possible to integrate there over $p_0$. So the unrenormalized
TDP (\ref{012}) can be presented in the following form,
\begin{eqnarray}
\Omega^{un} (M,\Delta)&=&
\frac{M^2}{4G_1}+\frac{\Delta^2}{4G_2}-
\frac{1}{4}\sum_{\eta =\pm}\int\frac{d^2p}{(2\pi)^2}\Big (|p^\eta_{01}|+|p^\eta_{02}|+|p^\eta_{03}|+|p^\eta_{04}|\Big ). \label{28}
\end{eqnarray}

\subsection{Renormalization and phase structure in the vacuum case: $\mu=0,\mu_5=0$}

First of all, let us obtain a finite, i.e. renormalized, expression for the TDP (\ref{28}) at $\mu=0$ and $\mu_5=0$, i.e. in vacuum. Since in this case a thermodynamic potential is usually called effective potential, we use for it the notation $V^{un} (M,\Delta)$. It follows from (\ref{012}) and (\ref{013}) that at $\mu=0$ and $\mu_5=0$ $V^{un} (M,\Delta)$ looks like
\begin{eqnarray}
V^{un} (M,\Delta)&=&
\frac{M^2}{4G_1}+\frac{\Delta^2}{4G_2}
+i\int\frac{d^3p}{(2\pi)^3}\ln\Big
[(p_0^2-E^2_+)(p_0^2-E^2_-)\Big ]\nonumber\\
&\equiv&\frac{M^2}{4G_1}+\frac{\Delta^2}{4G_2}-\int\frac{d^2p}{(2\pi)^2}\Big (\sqrt{|\vec p|^2+(M+\Delta)^2}+\sqrt{|\vec p|^2+(M-\Delta)^2}\Big ), \label{25}
\end{eqnarray}
where $E_\pm=\sqrt{|\vec p|^2+(M\pm\Delta)^2}$. (To obtain the
second line of this formula, we have integrated there over $p_0$
according to the relation (\ref{int}).) It is evident that
the effective potential (\ref{25}) is an ultraviolet divergent
quantity. So, we need to renormalize it. This procedure consists of
two steps: (i) First of all we need to regularize the divergent
integral in (\ref{25}), i.e. we suppose there that $|p_1|<\Lambda$,
$|p_2|<\Lambda$. (ii)  Second, we must suppose also that the bare
coupling constants $G_1$ and $G_2$ depend on the cutoff parameter
$\Lambda$ in such a way that in the limit $\Lambda\to\infty$ one
obtains a finite expression for the effective potential.

Before performing the steps (i) and (ii)  of the renormalization
procedure, it is useful to take into account the following
asymptotic expansion at $|\vec p|\to\infty$
\begin{eqnarray}
\sqrt{|\vec p|^2+(M+\Delta)^2}+\sqrt{|\vec p|^2+(M-\Delta)^2}=2|\vec p|+\frac{(M^2+\Delta^2)}{|\vec p|}+{\cal O}\big (1/|\vec p|^3\big ).
\label{19}
\end{eqnarray}
Then, after construction of the  regularized effective potential
$V^{reg} (M,\Delta)$ (see below), we use there the asymptotic
expansion (\ref{19}) and integrate it over $p_1$ and $p_2$
term-by-term. The result reads
\begin{eqnarray}
V^{reg} (M,\Delta)&\equiv&\frac{M^2}{4G_1}+\frac{\Delta^2}{4G_2}-\int_{-\Lambda}^{\Lambda}\frac{dp_1}{2\pi}\int_{-\Lambda}^{\Lambda}\frac{dp_2}{2\pi}\Big (\sqrt{|\vec p|^2+(M+\Delta)^2}+\sqrt{|\vec p|^2+(M-\Delta)^2}\Big )\nonumber\\
=M^2\left [\frac
1{4G_1}\right.&-&\left.\frac{2\Lambda\ln(1+\sqrt{2})}{\pi^2}\right
]+\Delta^2\left [\frac
1{4G_2}-\frac{2\Lambda\ln(1+\sqrt{2})}{\pi^2}\right
]-\frac{2\Lambda^3(\sqrt{2}+\ln(1+\sqrt{2}))}{3\pi^2}+{\cal
O}(\Lambda^0), \label{12}
\end{eqnarray}
where ${\cal O}(\Lambda^0)$ denotes an expression which is finite in
the limit $\Lambda\to \infty$.  Now, we should suppose that the bare
coupling constants $G_1$ and $G_2$ depend on the cutoff parameter
$\Lambda$ in such a way that in the limit $\Lambda\to\infty$ one obtains a finite expressions in the square brackets of (\ref{12}). Clearly, to fulfil this requirement it is sufficient to require that
 \begin{eqnarray}
\frac 1{4G_1}\equiv \frac
1{4G_1(\Lambda)}=\frac{2\Lambda\ln(1+\sqrt{2})}{\pi^2}+\frac{1}{2\pi g_1},
~~~\frac 1{4G_2}\equiv \frac
1{4G_2(\Lambda)}=\frac{2\Lambda\ln(1+\sqrt{2})}{\pi^2}+\frac{1}{2\pi g_2},
\label{13}
\end{eqnarray}
where $g_{1,2}$ are finite and $\Lambda$-independent model
parameters with dimensionality of inverse mass. Since bare couplings
$G_1$ and $G_2$ do not depend on a normalization point, the  same
property is also valid for $g_{1,2}$. Hence, taking into account in
(\ref{12}) the relations (\ref{13}) and ignoring there an infinite
$M$- and $\Delta$-independent constant, one obtains the following
{\it renormalized}, i.e. finite, expression $V^{ren}(M,\Delta)$ for
the effective potential,
\begin{eqnarray}
V^{ren}(M,\Delta)&=&\lim_{\Lambda\to\infty}
\left\{V^{reg}(M,\Delta)\Big |_{G_1= G_1(\Lambda),G_2=
G_2(\Lambda)}+\frac{2\Lambda^3(\sqrt{2}+\ln(1+\sqrt{2}))}{3\pi^2}\right\}.\label{14}
\end{eqnarray}
It should also be mentioned that the effective potential (\ref{14}) is a
renormalization group invariant quantity.

The fact that it is possible to renormalize the effective potential
of the initial model (1) in the leading order of the large
$N$-expansion is the reflection of a more general property of
(2+1)-dimensional theories with four-fermion interactions. Indeed,
it is well known that in the framework of  the ''naive''
perturbation theory (over coupling constants) these models are not
renormalizable. However, as it was proved in \cite{28},
in the framework of nonperturbative large $N$-technique these models
are renormalizable in each order of the $1/N$-expansion.

The ${\cal O}(\Lambda^0)$ term in
(\ref{12}) can be calculated explicitly, so we have for the
renormalized effective potential $V^{ren}(M,\Delta)$ (\ref{14}) the following expression
\begin{eqnarray}
V^{ren}(M,\Delta)=
\frac{M^2}{2\pi g_1}+\frac{\Delta^2}{2\pi g_2}+\frac{(M+\Delta)^{3}}{6\pi}+\frac{|M-\Delta|^{3}}{6\pi}.\label{15}
\end{eqnarray}
The global minimum point $(M_0,\Delta_0)$, where $M_0=\vev{\sigma(x)}$ and $\Delta_0=\vev{\Delta(x)}$, of this function was already investigated in
\cite{Zhukovsky:2000yd}, although in the framework of another
(2+1)-dimensional GN model. So, we present at once the phase structure
of the initial model (1) at $\mu=0$ and $\mu_5=0$ (see Fig. 1).

In Fig. 1 the phase portrait of the model is depicted depending on
the values of the free model parameters $g_1$ and $g_2$. There the
plane $(g_1,g_2)$ is  divided into several areas.  In each area one
of the phases I, II or III is implemented. In the phase I, i.e. at
$g_1>0$ and $g_2>0$, the global minimum of the effective potential
$V(M,\Delta)$ is arranged at the origin. So in this case we have
$M_0=0$ and $\Delta_0=0$. As a result, in the phase I both continuous symmetries, chiral $U_{\gamma^5}(1)$ and electromagnetic $U_V(1)$, remain intact and fermions are massless. Due to this reason the phase I is called symmetric. In the phase II, which is allowed only for $g_1<0$, at the global minimum point the relations $M_0=-1/g_1$ and
$\Delta_0=0$ are valid. So in this phase chiral $U_{\gamma^5}(1)$ symmetry is spontaneously broken down and fermions acquire dynamically the mass
$M_0$. Finally, in the superconducting phase III, where $g_2<0$, we
have spontaneous breaking of the $U_V(1)$ symmetry. This phase corresponds to the following values of the gaps: $M_0=0$, $\Delta_0=-1/g_2$.

Note also that if $g_1=g_2\equiv g$ and, in addition, $g<0$ (it is
just the line {\it l} in Fig. 1), then the effective potential (\ref{15})
has two equivalent global minima. The first one, the point
$(M_0=-1/g,\Delta_0=0)$, corresponds to a phase with chiral symmetry
breaking. The second one, i.e. the point $(M_0=0,\Delta_0=-1/g)$,
corresponds to superconductivity.

Clearly, if the cutoff parameter $\Lambda$ is fixed, then the phase
structure of the model can be described in terms of bare coupling
constants $G_1, G_2$ instead of finite quantities $g_1, g_2$.
Indeed, let us first introduce a critical value of the couplings,
$G_c=\frac{\pi^2}{8\Lambda\ln(1+\sqrt{2})}$. Then, as it follows
from Fig. 1 and (\ref{13}), at $G_1<G_c$ and $G_2<G_c$ the symmetric
phase I of the model is located. If $G_1>G_c$, $G_2<G_c$ ($G_1<G_c$,
$G_2>G_c$), then the chiral symmetry broken phase II (the
superconducting phase III) is realized. Finally, let us suppose that
both $G_1>G_c$ and $G_2>G_c$. In this case at $G_1>G_2$ ($G_1<G_2$)
we have again the chiral symmetry broken phase II (the
superconducting phase III).

Now, few comments about the nature of the superconductivity (SC) and
chiral symmetry breaking (CSB) are in order. In the framework of the
model (1) there are two well-known mechanisms for appearing of these
phenomena, (i) dynamical symmetry breaking, which occurs at strong
couplings, and (ii) Cooper instability of Fermi surface, which takes
place, in contrast to the case (i), at weak couplings. If $\mu=0$
and $\mu_5=0$, then the initial model (1) describes effectively the
undoped regime of high temperature cuprate superconductors, etc
(see, e.g., the papers \cite{marino}, where the phase structure,
similar to the phase diagram of our Fig. 1, was obtained). In this case Fermi surface is absent, so both CSB
and SC phases appear dynamically in the system at a {\it rather
strong attraction} in the fermion-antifermion or difermion channels,
i.e. at $G_1>G_c$ and $G_1>G_2$ or at $G_2>G_c$ and $G_2>G_1$,
respectively. Hence, at zero values of chemical potentials only the mechanism (i) for generating both SC and CSB is realized in our model.
\footnote{The mechanism of dynamical symmetry breaking at $\mu=0$ is
well known since the papers by Nambu and Jona-Lasinio [3]. It is
also valid in the framework of other (3+1)-dimensional 4F models
with chiral and color superconducting channels of interaction at
$\mu=0$ \cite{van}. In \cite{van} the bare coupling constants
$G_1$ and $G_2$ are free model parameters. Moreover, strong constraints were obtained there on $G_1$ and $G_2$, at which color SC or CSB is generated dynamically in the models.}

The second way, i.e. the mechanism (ii), to break spontaneously a
symmetry can be realized only at nonzero chemical potentials. For
example, if the doping or impurities are present in the system, then one
must introduce a finite chemical potential $\mu$. In this case the Fermi
surface can be created. As a result, if there is an {\it arbitrary small
attraction} in the SC channel, i.e. even at $G_2<G_c$, there appears
the so-called Cooper instability of the Fermi surface, which
destroys the normal ground state in favor of superconductivity.
(Below, when discussing the phase portrait of Fig. 5 (see Sect \ref{IVB}), we will see that at $\mu_5=0$ and $\mu>0$ both (i) and (ii)
mechanisms take part in forming the phase structure.) If $\mu=0$ but $\mu_5\ne 0$, then CSB can also be generated in the model due to the mechanism (ii) at arbitrary small values of $G_1$ (see in \cite{Braguta}). This fact supports the duality correspondence between SC and CSB in the model (1). 
\begin{figure}
\includegraphics[width=0.45\textwidth]{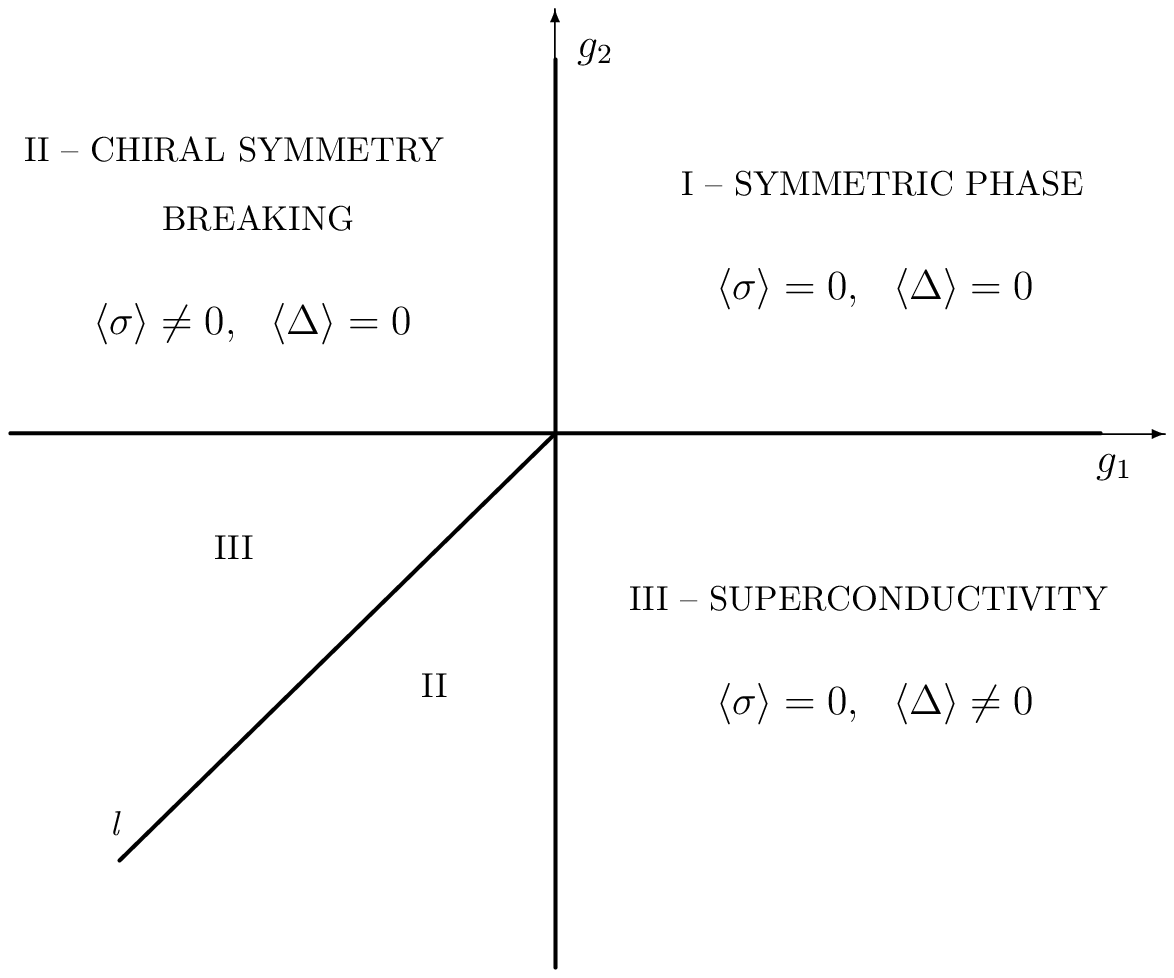}
\hfill
\includegraphics[width=0.45\textwidth]{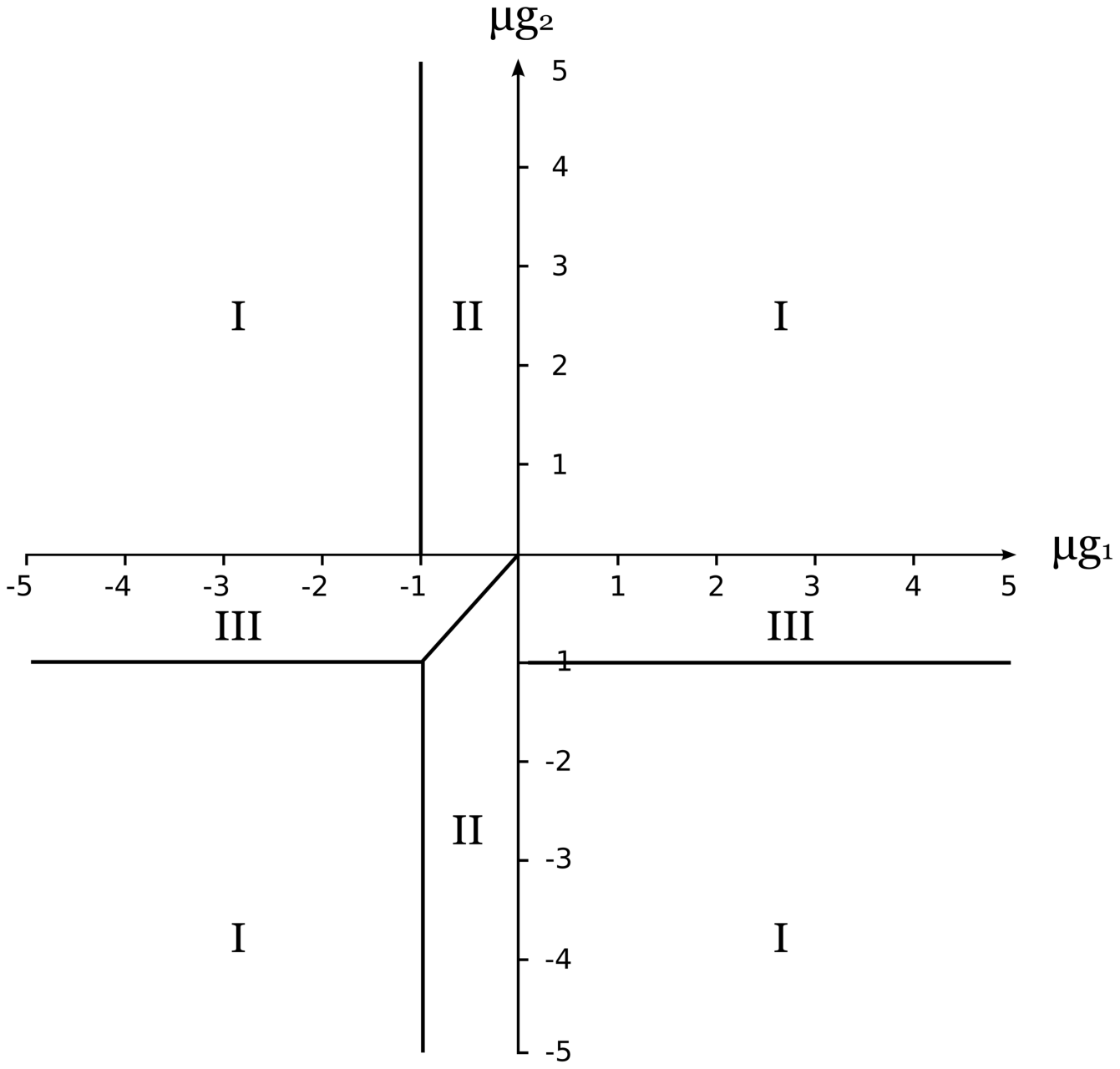}
\\
\parbox[t]{0.45\textwidth}{
\caption{The $(g_1,g_2)$-phase portrait of the model at $\mu=0$ and
$\mu_5=0$. The notations I, II and III mean the symmetric, the
chiral symmetry breaking (CSB) and the superconducting (SC) phases,
respectively. At $g_{1,2}<0$ the line {\it l} is defined by the
relation {\it l} $\equiv\{(g_1,g_2):g_1=g_2\}$.
 } } \hfill
\parbox[t]{0.45\textwidth}{
\caption{The $(g_1,g_2)$-phase portrait of the model at  fixed
chemical potentials, such that $\mu=\mu_5$. The notations I, II and
III are the same as in Fig. 1.}}
\end{figure}

\subsection{Renormalization of the TDP (\ref{28}) in the
general case}

To renormalize the TDP (\ref{28}) in the most general case, it is necessary to imagine how the integrand of (\ref{28}), i.e. the quantity $\sum_{\eta =\pm}\sum_{i=1}^4|p^\eta_{0i}|$, behaves at asymptotically high values of $|\vec p|$. The properties of the roots $p^\eta_{0i}$ ($i=1,...,4$) of the polynomial $P_\eta(p_0)$ appearing in (\ref{012}) can be obtained both numerically and analytically using the methods of Appendix B of \cite{ekkz}. So it is easy to show that at $|\vec p|\to\infty$
\begin{eqnarray}
\frac 14\sum_{\eta =\pm}\sum_{i=1}^4|p^\eta_{0i}|=2|\vec p|+\frac{(M^2+\Delta^2)}{|\vec p|}+{\cal O}\big (1/|\vec p|^3\big ). \label{016}
\end{eqnarray}
Since the asymptotic expansions (\ref{016}) and (\ref{19}) coincide,
one can subtract from the integrand of (\ref{28}) the expression
$4(E_++E_-)$, thereby obtaining a convergent integral. Taking into
account this circumstance, we have identically
\begin{eqnarray}
\Omega^{un} (M,\Delta)=V^{un}(M,\Delta)-\widetilde\Omega(M,\Delta),
\label{017}
\end{eqnarray}
where $V^{un}(M,\Delta)$ is the effective potential (\ref{25}) of the model in vacuum, i.e. at $\mu=0$ and $\mu_5=0$, and
\begin{eqnarray}
\widetilde\Omega(M,\Delta)&=&\frac{1}{4}\int\frac{d^2p}{(2\pi)^2}\Big
[\sum_{\eta =\pm}\sum_{i=1}^4|p^\eta_{0i}|
-4\sqrt{|\vec p|^2+(M+\Delta)^2}-4\sqrt{|\vec p|^2+(M-\Delta)^2}\Big ].\label{018}
\end{eqnarray}
It is clear that the term $\widetilde\Omega(M,\Delta)$ in (\ref{017})
is a convergent integral and all ultraviolet divergences of the TDP
(\ref{28})-(\ref{017}) are located in the term
$V^{un}(M,\Delta)$. Hence, we can renormalize only the first term of
(\ref{017}) in the way of the previous subsection and obtain finally:
\begin{eqnarray}
\Omega^{ren} (M,\Delta)=V^{ren}(M,\Delta)-\widetilde\Omega(M,\Delta),
\label{20}
\end{eqnarray}
where $V^{ren}(M,\Delta)$ and  $\widetilde\Omega(M,\Delta)$ are
presented in (\ref{15}) and (\ref{018}), correspondingly.

It was already mentioned above that using the method presented in
Appendix B of the paper \cite{ekkz}, the roots $p^\eta_{0i}$ can be calculated numerically (at fixed values of $|\vec p|, M,\Delta$, etc). As a result, it is also possible to study numerically the whole TDP (\ref{20}). Carrying out this procedure, we have seen that the global minimum of the TDP $\Omega^{ren} (M,\Delta)$ is always at the point of the form $(M\ge 0,\Delta=0)$ or $(M= 0,\Delta\ge 0)$. So, to get a more detailed information about the phase structure of the model, it is sufficient to investigate the reductions of the TDP (\ref{20}) on the $M$- and $\Delta$ axes \footnote{This procedure can be easily performed since the reduction of the roots $p^\eta_{0i}$ on the $M$- and $\Delta$ axes is known (see (\ref{26}) and (\ref{27})).} , i.e. to consider the functions
\begin{eqnarray}
F_1(M)&\equiv&\Omega^{ren}(M,\Delta=0)=V_1(M)-\frac{1}{2}\int\frac{d^2p}{(2\pi)^2} \sum_{\eta
=\pm}\Big [\mu+{\cal E}_{\eta}+|\mu-{\cal E}_{\eta}|-2\sqrt{|\vec p|^2+M^2}\Big
]\nonumber\\
&=&V_1(M)-\int\frac{d^2p}{(2\pi)^2} \sum_{\eta =\pm}\Big
[{\cal E}_{\eta}-\sqrt{|\vec p|^2+M^2}+(\mu-{\cal E}_{\eta})\theta(\mu-{\cal E}_{\eta})\Big ],
\label{31}\\
F_2(\Delta)&\equiv&\Omega^{ren}(M=0,\Delta)=
V_2(\Delta)-\int\frac{d^2p}{(2\pi)^2} \sum_{\eta =\pm}\Big
[\sqrt{\Delta^2+(|\vec p|+\eta\mu)^2}-\sqrt{|\vec p|^2+M^2}\nonumber\\&+&\left (\mu_5-\sqrt{\Delta^2+(|\vec p|+\eta\mu)^2}\right )\theta\left (\mu_5-\sqrt{\Delta^2+(|\vec p|+\eta\mu)^2}\right )\Big ],
\label{32}
\end{eqnarray}
respectively, where ${\cal E}_{\eta}=\sqrt{M^2+(|\vec
p|+\eta\mu_5)^2}$ and $3\pi V_i(x)=x^3+3x^2/(2g_i)$ is the reduction
of the vacuum effective potential (\ref{15}) on the $M$ (in this
case $i=1$, $x=M$) or $\Delta$ (in this case $i=2$, $x=\Delta$)
axes. Moreover, to obtain the second line in (\ref{31}) we use the
evident relations, $|x|=x\theta (x)-x\theta (-x)$ and $1=\theta
(x)+\theta (-x)$. After tedious but straightforward calculations,
the TDP (\ref{31}) can be presented in the following form (see
Appendix \ref{ApC}):
\begin{eqnarray}
F_1(M)&=&\frac{M^2}{2\pi g_1}+\frac{(\mu_5^2+M^2)^{3/2}}{3\pi}-\frac{\theta\left (\mu-\sqrt{M^2+\mu_5^2}\right )}{6\pi}\left [\mu^3-3\mu(M^2-\mu_5^2)+2(\mu_5^2+M^2)^{3/2}\right ]\nonumber\\
&&-\frac{\theta\left (\sqrt{M^2+\mu_5^2}-\mu\right )}{2\pi}\left [\mu_5^2\sqrt{\mu_5^2+M^2}+\mu_5 M^2\ln\left (\frac{\mu_5+\sqrt{\mu_5^2+M^2}}{M}\right )\right ]\nonumber\\
&&-\frac{\theta\left (\mu-M\right )\theta\left (\sqrt{M^2+\mu_5^2}-\mu\right )}{2\pi}\left [\mu_5\mu\sqrt{\mu^2-M^2}-\mu_5 M^2\ln\left (\frac{\mu+\sqrt{\mu^2-M^2}}{M}\right )\right ]\label{C18}
\end{eqnarray}
(at $\mu_5=0$ this expression coincides with the corresponding TDP (22) from the paper \cite{kzz}). Moreover, comparing the expressions (\ref{31}) and (\ref{32}), we see that the quantity $\Omega^{ren} (M=0,\Delta)$ can be easily obtained from the TDP $\Omega^{ren} (M,\Delta=0)$ by the substitutions $M\to\Delta$, $g_1\to g_2$ and $\mu\leftrightarrow\mu_5$, i.e.
\begin{eqnarray}
F_2(\Delta)=F_1(\Delta)\Big |_{g_1\to g_2,\mu\leftrightarrow\mu_5}.
\label{35}
\end{eqnarray}
Clearly, the connection (\ref{35}) between the reductions  of the
TDP $\Omega^{ren} (M,\Delta)$ on the $M$- and $\Delta$ axes is the
consequence of the duality property (\ref{16}) of the model.

Next, in order to find the phase structure of  the model, we will
determine the global minimum points of the TDPs $F_1(M)$ (\ref{C18})
and $F_2(\Delta)$ (\ref{35}) and then compare the minimum values of
these functions vs external parameters $\mu,\mu_5,g_1,g_2$.

\section{Phase structure: the role of the
duality invariance of the TDP}

Suppose now that at some fixed particular values of the model
parameters, i.e. at $g_1=A$, $g_2=B$ and $\mu=\alpha$,
$\mu_5=\beta$, the global minimum of the TDP (\ref{20}) lies at the
point, e.g., $(M=M_0,\Delta=0)$. It means that for such fixed values
of the parameters the chiral symmetry breaking (CSB) phase is
realized. Then it follows from the duality invariance of the
unrenormalized TDP with respect to the transformation ${\cal D}$
(\ref{16}) that the permutation of the coupling constant and
chemical potential values \footnote{It is evident that the duality
transformation ${\cal D}$ for the renormalized TDP (\ref{20}) means
$g_1\longleftrightarrow g_2$, $M\longleftrightarrow \Delta$,
$\mu\longleftrightarrow\mu_5$, under which this TDP is invariant.}
(i.e. at $g_1=B$, $g_2=A$ and $\mu=\beta$,  $\mu_5=\alpha$) moves
the global minimum of the TDP $\Omega^{ren}(M,\Delta)$ to the point
$(M=0,\Delta=M_0)$, and the superconducting (SC) phase is originated
(and vice versa). This is the so-called duality correspondence
between CSB and SC phases in the framework of the model under
consideration. Hence, the knowledge of a phase of the model (1) at
some fixed values of external free model parameters
$g_1,g_2,\mu,\mu_5$ is sufficient  to understand what phase is
realized at rearranged values of external parameters,
$g_1\leftrightarrow g_2,~~\mu\leftrightarrow\mu_5$. Moreover, we
would like to emphasize once again that there exists an equality of
the order parameters (condensates), which characterizes both the
initial phase and the phase corresponding to rearranged external
parameters. In other words the chiral condensate of the CSB phase at
fixed $g_1,g_2,\mu,\mu_5$ is equal to the SC condensate of the phase
at $g_1\leftrightarrow g_2,~~\mu\leftrightarrow\mu_5$ (and vice
versa).

The duality transformation (\ref{16}) of the  TDP can also be
applied to the arbitrary phase portrait of the model (see below). In
particular, it is clear that if we have a most general phase
portrait, i.e. the correspondence between any point
$(g_1,g_2,\mu,\mu_5)$ of the four-dimensional space of external
parameters and possible model phases (CSB, SC and symmetric phase),
then under the duality transformation ($g_1\leftrightarrow g_2$,
$\mu\leftrightarrow\mu_5$, CSB$\leftrightarrow$SC) this phase
portrait is mapped to itself, i.e. the most general phase portrait
is selfdual. In practice, usually, there are constraints on the
model parameters. As a result, if the constraint is dually
(non)invariant, then the phase portrait is also a dually
(non)invariant.

Below, we will use the dual symmetry of the TDP in order to explain
and construct the phase structure of the model in different particular
cases.
\begin{figure}
\includegraphics[width=0.45\textwidth]{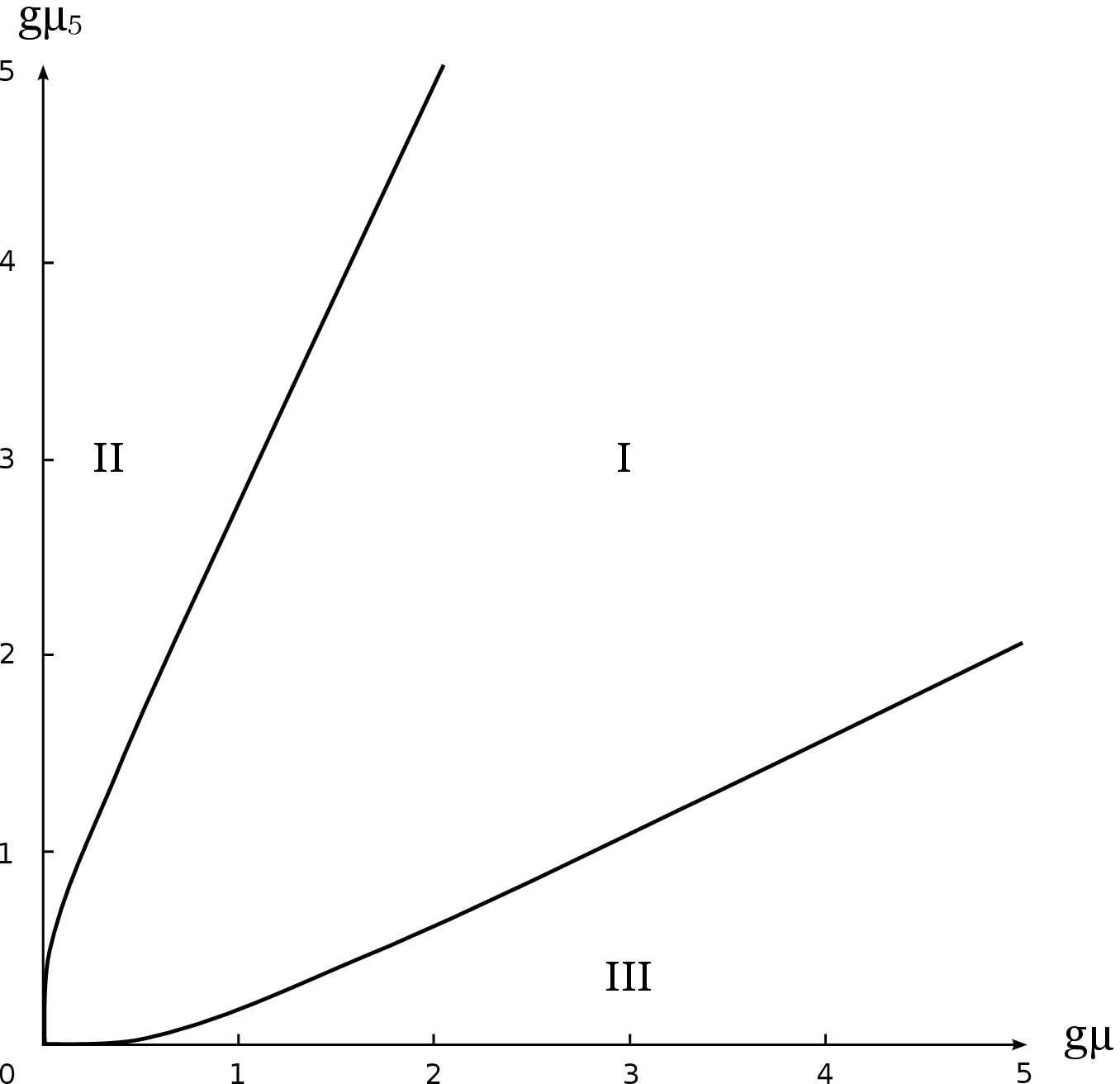}
\hfill
\includegraphics[width=0.45\textwidth]{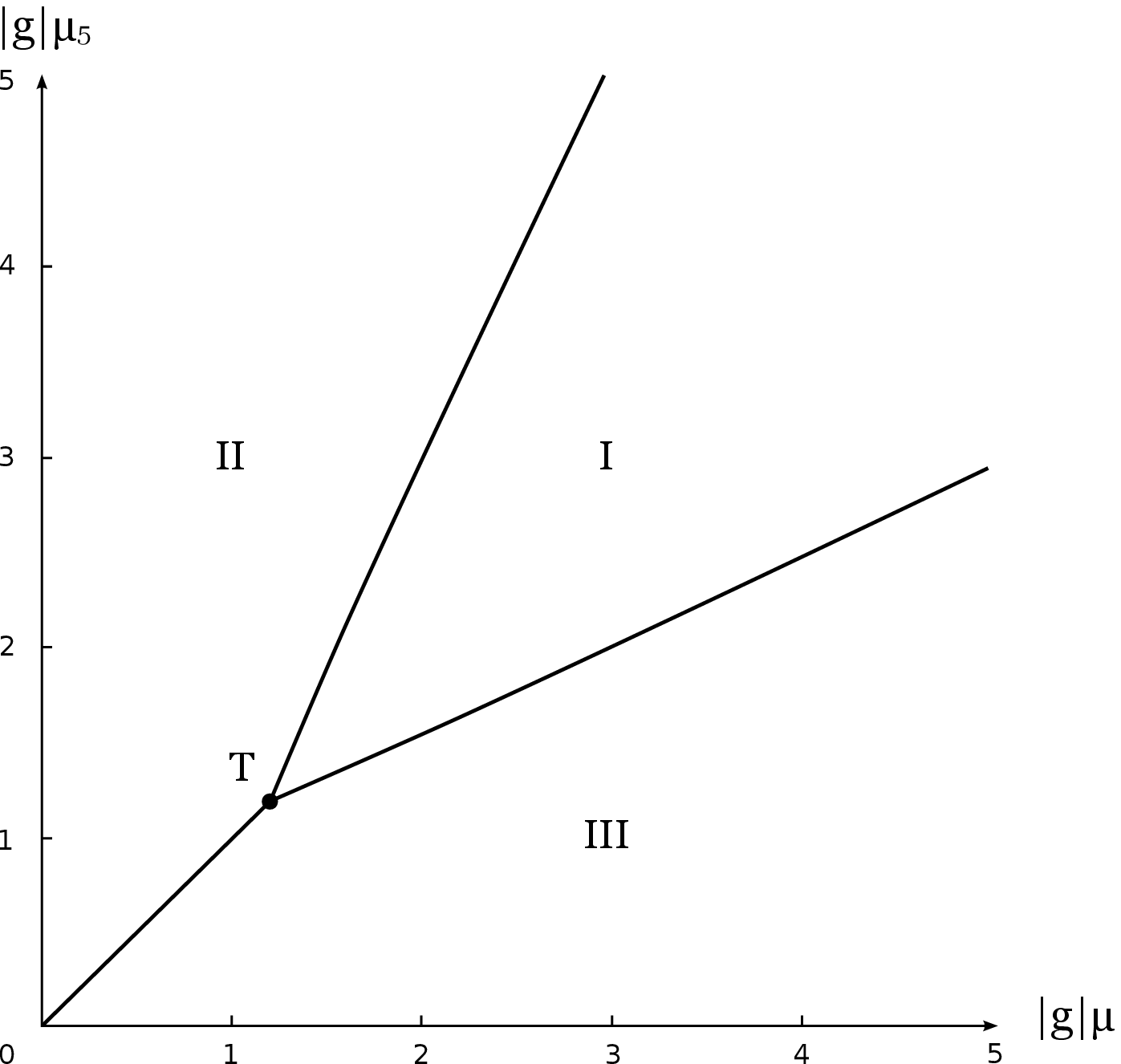}
\\
\parbox[t]{0.45\textwidth}{
\caption{The $(\mu,\mu_5)$-phase portrait of the model at fixed coupling constants, such that $g_1=g_2\equiv g>0$.  The notations I, II and III are the same as in Fig. 1. } } \hfill
\parbox[t]{0.45\textwidth}{
\caption{The $(\mu,\mu_5)$-phase portrait of the model at fixed
coupling constants, such that $g_1=g_2\equiv g<0$. The notations I,
II and III are the same as in Fig. 1. The letter $T$ denotes a triple point.}}
\end{figure}

\subsection{Selfdual phase portraits}

First of all, taking into account the duality correspondence between
CSB and SC, let us determine its  characteristic features and then
find (analytically and numerically) the phase structure of the model
(1) in two particular cases, (i) $\mu=\mu_5$  and (ii) $g_1=g_2$. It
is evident that both of these constraints, (i) and (ii), are
invariant with respect to the duality transformation
$g_1\leftrightarrow g_2$, $\mu\leftrightarrow\mu_5$. Hence, the
corresponding phase portraits should be selfdual (a more detailed
explanation of this fact is presented in the next subsection). In
the case (i) we want to study the $(g_1,g_2)$ phase portrait,
whereas in the case (ii) -- the $(\mu,\mu_5)$  phase portrait. It
follows from the duality correspondence that in the case (i) the
region of the CSB phase is a mirror image of the SC phase with
respect to the line  $g_1=g_2$ of the $(g_1,g_2)$ plane. However, in
the case (ii) the regions of these phases are arranged mirror
symmetrically  with respect to the line $\mu=\mu_5$ of the plane
$(\mu,\mu_5)$.

Of course, the specific form of  the CSB and SC areas depends on the
values of the external parameters. For example, at $\mu=\mu_5=0$ the
phase structure is presented in Fig. 1. (Note that in order to
obtain this phase portrait it is sufficient to investigate the
behavior of the global minimum point of the TDP $V^{ren}(M,\Delta)$
(\ref{15}) vs the coupling constants $g_1$ and $g_2$.) However, at
fixed $\mu=\mu_5>0$ the phase structure is presented in Fig. 2,
where one can see the mirror symmetrical arrangement of the phases
II and III with respect to the line $g_1=g_2$.

In Figs 3, 4 the $(\mu,\mu_5)$-phase portraits of the model (1) are
presented at fixed $g_1=g_2\equiv g>0$ and fixed $g_1=g_2\equiv
g<0$, respectively. The CSB and SC phases in these figures are
arranged, as defined above, symmetrically with respect to the line
$\mu=\mu_5$ of the $(\mu,\mu_5)$ plane. If $g>0$, then at the points
of the boundary of the phase I in Fig. 3, the second order phase
transition takes place. In this case it is possible to give
exact analytical expressions for the boundaries of the region I in Fig. 3.
Indeed, the global minimum point $M_0$ of the TDP (\ref{C18}) is
defined by the equation $\partial F_1(M)/\partial M\equiv Mf(M)=0$.
Inside the region II of Fig. 3 we have $M_0>0$ and $f(M_0)=0$.
However, on the boundary between phases I and II one can see that
$M_0=0$ and, moreover, that $f(0)=0$. It is the last relation that
defines the boundary equation between  I and II phases. This
equation looks like
\begin{eqnarray}
\mu=\mu_5\exp \left (-\frac{1}{g\mu_5}-1\right ).
\label{36}
\end{eqnarray}
In a similar way it is possible to  find the equation for the
boundary between I and III phases of Fig. 3:
\begin{eqnarray}
\mu_5=\mu\exp \left (-\frac{1}{g\mu}-1\right ).
\label{37}
\end{eqnarray}
In contrast, on the boundary of the  phase I of Fig. 4, a phase
transition of the first order takes place. Since on the boundary
between II and III phases of this figure we also have a first order
phase transitions, it is clear that the point T of the phase
portrait Fig. 4 is a so-called triple point, i.e. the point where
three different phases, I, II and III, coexist. The triple point T
of Fig. 4 corresponds to the chemical potential values
$\mu_T=\mu_{5T}\approx 1.2/|g|$.

\subsection{Nonselfdual phase portraits and their dual transformations}
\subsubsection{The cases $\mu_5=0$, $\mu\ne 0$ and $\mu=0$, $\mu_5\ne 0$}
\label{IVB}

In these two particular cases the  restrictions on the external
model parameters are already not dually invariant, so the
corresponding $(g_1,g_2)$-phase portraits are not selfdual. However,
as we shall see, each of these two phase portraits is a dual mapping
of another one.

The $(g_1,g_2)$-phase structure of the model at arbitrary fixed
nonzero value of $\mu$ (of $\mu_5$) and at $\mu_5=0$ (at $\mu=0$) is
presented in Fig. 5 (in Fig. 6). Let us first compare Fig. 1 and Fig. 5. It is easy to see that if $g_1>0$, $g_2>0$ and $\mu=\mu_5=0$, then the
system is in the symmetric phase I (see Fig. 1). However, an
arbitrary small nonzero value of the chemical potential $\mu$ induces
(at $\mu_5=0$) in this case the superconducting phase III (see Fig. 5
as well as Fig. 3 for the particular case $g_1=g_2>0$). 
Note that there is a very simple explanation of this fact, which is based on the symmetry breaking mechanism (ii) (see the end of the section III A), i.e. on the Cooper instability of a Fermi surface. Indeed, at  $g_1>0$, $g_2>0$ and $\mu>0$ we have a
nonzero particle density (see, e.g., in \cite{kzz}), so there is a Fermi sea of particles with energies less or equal to $\mu$ (Fermi surface). 
Evidently, in this case there is no energy cost for creating a pair
of particles with opposite momenta just over the Fermi surface.
Then, due to an arbitrary weak attraction between these particles
($0<G_2<G_c$), the Cooper pair is formed and $U_V(1)$ symmetry is
spontaneously broken, as a result of Bose--Einstein condensation of
Cooper pairs. Since in the energy spectrum of fermions the gap
$\Delta\ne 0$ appears, rather small external forces
are not able to destroy the superconducting condensate and it is a
stable one.

If $g_1>0$, $g_2<0$, then we have a rather strong interaction in the SC channel, i.e. $G_2>G_c$ and $G_1<G_c$. As a result, in this case both at $\mu=0$ and $\mu>0$ the dynamical symmetry breaking mechnism (i) (see the end of the section III A) plays a decisive role in the forming of SC phenomenon.  

Now suppose
that $g_1< 0$ and $g_2\ne 0$, i.e. $G_1>G_c$. (In particular, the point $(g_1,g_2)$
might belong to the  phase II of Fig. 1, where chiral symmetry is
spontaneously broken at $\mu=0$, $\mu_5=0$.) Then it follows from
Fig. 5 that at $\mu_5=0$ there is a sufficiently high critical value
$\mu_c\ge 0$ of the chemical potential $\mu$, such that at $\mu>\mu_c$
the point $(\mu g_1,\mu g_2)$ will certainly be in one of the regions
III of Fig. 5. Hence, the growth (at $\mu_5=0$) of the chemical
potential $\mu$ leads to the appearance of the SC phase in the model
(1) at arbitrary $g_1\ne 0$ and $g_2\ne 0$. \footnote{This
was the main result of the paper \cite{kzz}, where the competition between CSB and SC was studied in the (2+1)-dimensional 4F model at $\mu\ne 0$, $\mu_5=0$ without duality correspondence.} Analysing in the same manner Fig. 6, we see that in the opposite case when $\mu=0$ but $\mu_5>0$, the growth (at $\mu=0$) of the chiral chemical potential $\mu_5$ leads to the appearance of the CSB phase in the model (1) at arbitrary fixed $g_1\ne 0$ and $g_2\ne 0$.

Indeed, these two properties of the model (1), i.e. (A) the
appearance of the SC at $\mu_5=0$ and at growing values of $\mu$ as
well as (B) the appearance of the CSB at $\mu=0$ and at growing
values of $\mu_5$ (in these two cases it is supposed that $g_1\ne 0$
and $g_2\ne 0$ are fixed), are dually connected (or dually
conjugated). To understand this fact, it is necessary to emphasize
once again that (A) follows from the phase portrait of Fig. 5,
whereas (B) is the consequence of the phase portrait presented in
Fig. 6. Now we will demonstrate that under the dual transformation
${\cal D}$ ($g_1\leftrightarrow g_2$, $\mu\leftrightarrow\mu_5$,
CSB$\leftrightarrow$SC) the phase portrait of Fig. 5 is transformed
into the phase portrait of Fig. 6 (and vice versa).

Note that the application of ${\cal D}$ to Fig. 5 can  be divided
into three more simple steps. (i) First, under permutation
$M\leftrightarrow\Delta$ in (\ref{16}) we have renaming of the
phases, i.e. II$\leftrightarrow$III. (For example, in this case the
global minimum point of the CSB phase, i.e. the point $(M_0,0)$, is
transformed into the point $(0,M_0)$ and, as a result, the CSB phase
is transformed into the SC phase.) (ii) Second, performing the
$\mu\leftrightarrow\mu_5$ and $g_1\leftrightarrow g_2$
transformations in Fig. 5, we rename the coordinate axes of the
figure (after the duality transformation we have along  the vertical
axis the quantity $\mu_5 g_1$ and along the horizontal one the
quantity $\mu_5 g_2$) and change its caption or external conditions,
at which the phase portrait is obtained. (In our case the
caption {\bf $\mu_5=0$ with $\mu$ being an arbitrary fixed quantity} for
Fig. 5 is replaced by {\bf $\mu=0$ with $\mu_5$ being an arbitrary
fixed quantity}, i.e. we obtain the caption for Fig. 6.)  (iii)
Finally, it is necessary to direct the axis corresponding to the
quantity $\mu_5 g_1$ (to the quantity $\mu_5 g_2$) horizontally
(vertically). As a result of this dual transformation of the Fig. 5
we obtain just Fig. 6.
\begin{figure}
\includegraphics[width=0.45\textwidth]{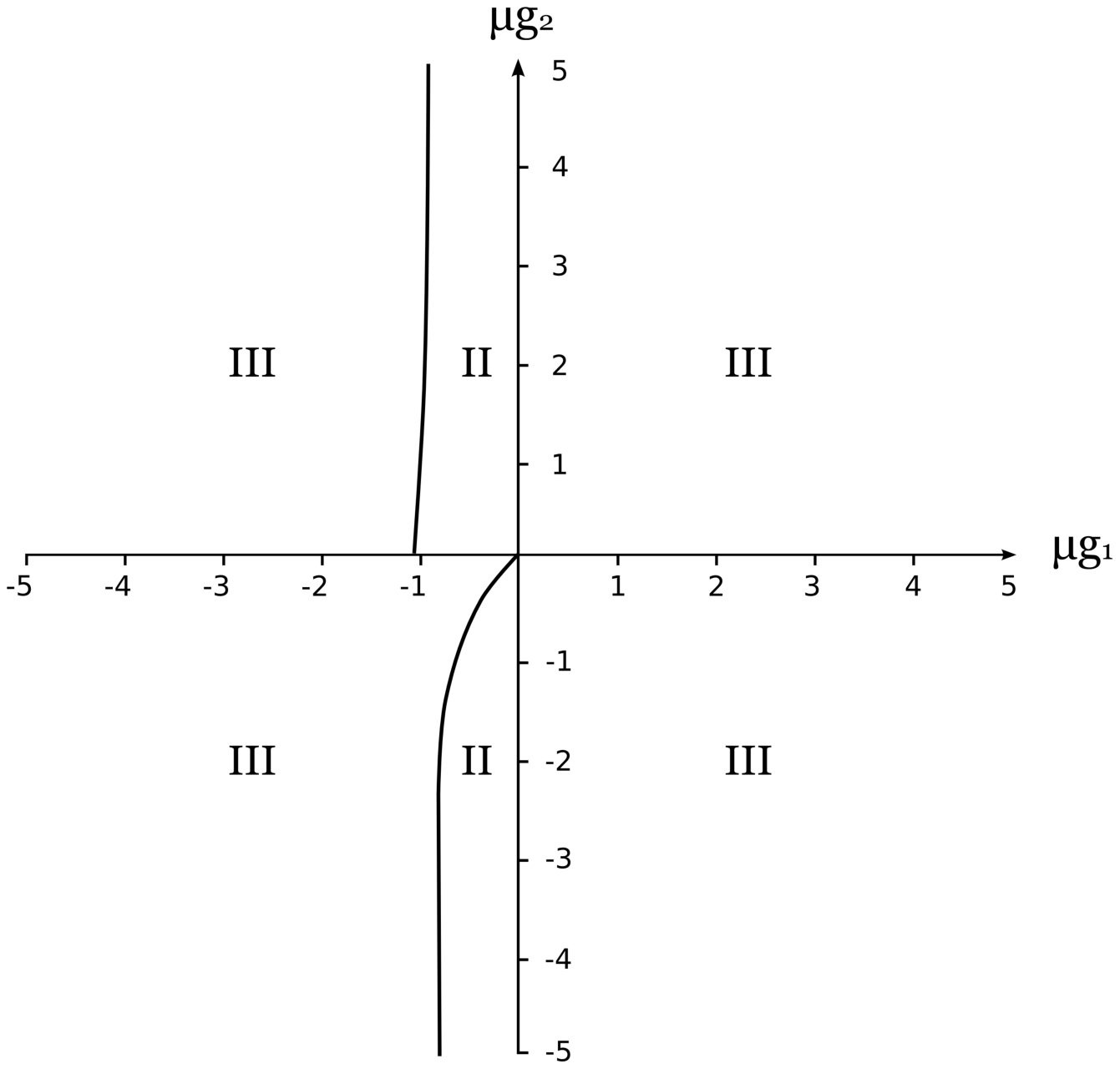}
\hfill
\includegraphics[width=0.45\textwidth]{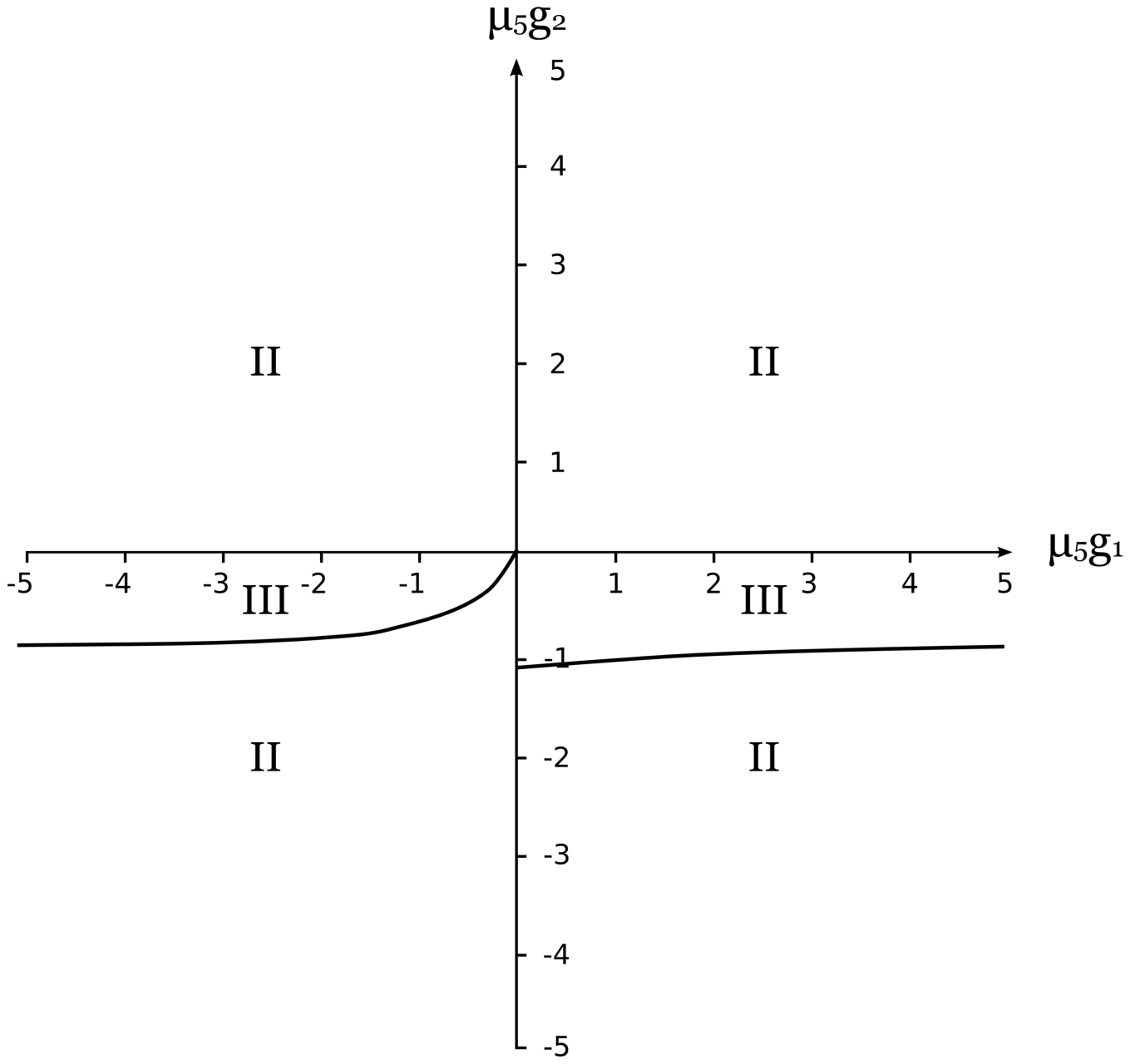}
\\
\parbox[t]{0.45\textwidth}{
\caption{The $(g_1,g_2)$-phase portrait  of the model at arbitrary
fixed nonzero value $\mu$ and at $\mu_5=0$. The notations I, II and
III are the same as in Fig. 1. } } \hfill
\parbox[t]{0.45\textwidth}{
\caption{The $(g_1,g_2)$-phase portrait  of the model at arbitrary
fixed nonzero value $\mu_5$ and at $\mu=0$. The notations I, II and
III are the same as in Fig. 1.}}
\end{figure}

In a similar way it is possible to apply the duality transformation
${\cal D}$ to each of the phase portraits presented in Figs. 1--4.
As a result, we see that these phase structures are selfdual. In the
next subsubsections we obtain the phase structure of the model in other
more general nonselfdual cases, i.e. the $(g_1,g_2)$-phase portraits
at arbitrary fixed values of chemical potentials and the
$(\mu,\mu_5)$-phase portraits at arbitrary fixed  values of the
coupling constants.
\begin{figure}
\includegraphics[width=0.45\textwidth]{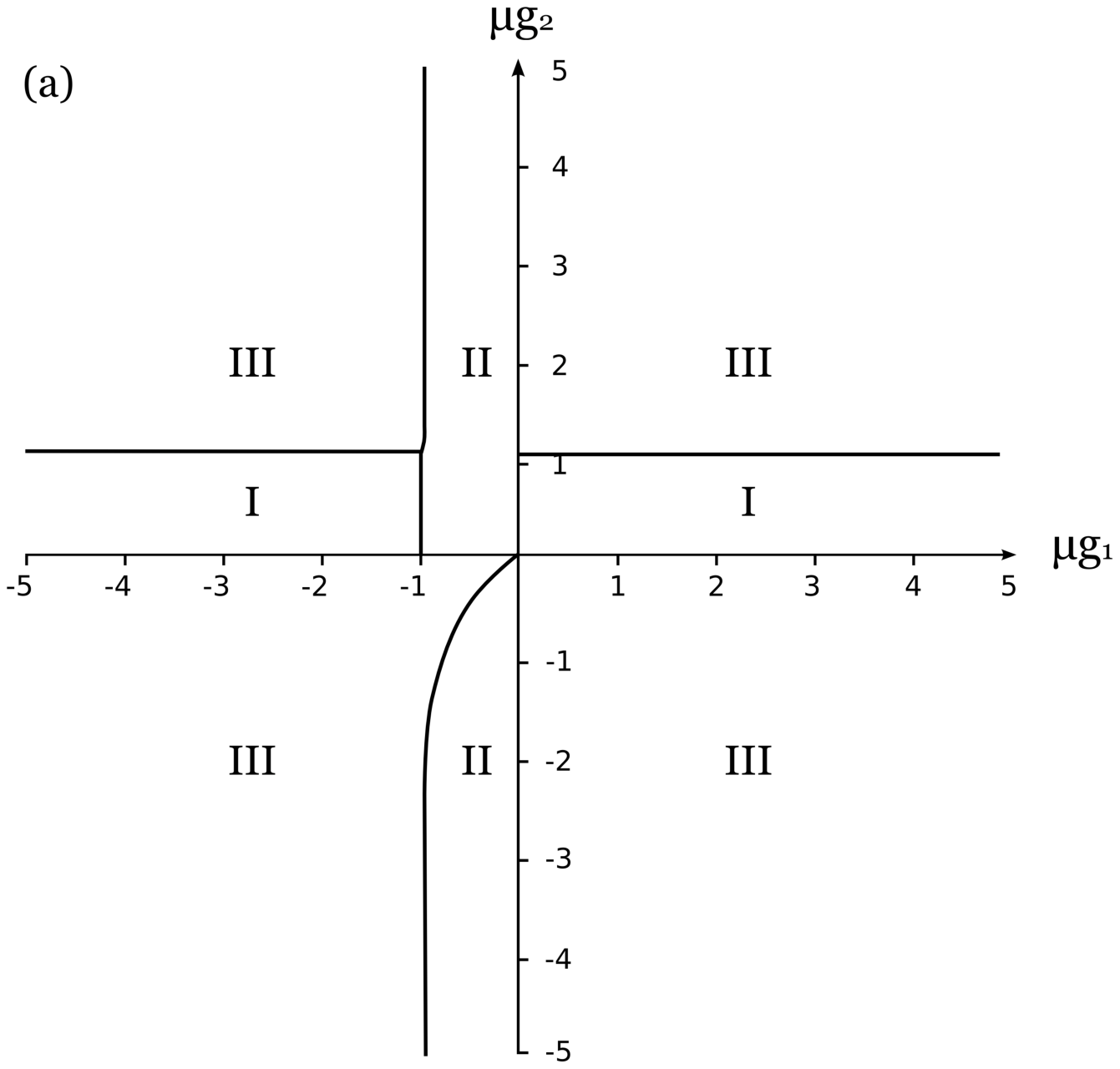}
\hfill
\includegraphics[width=0.45\textwidth]{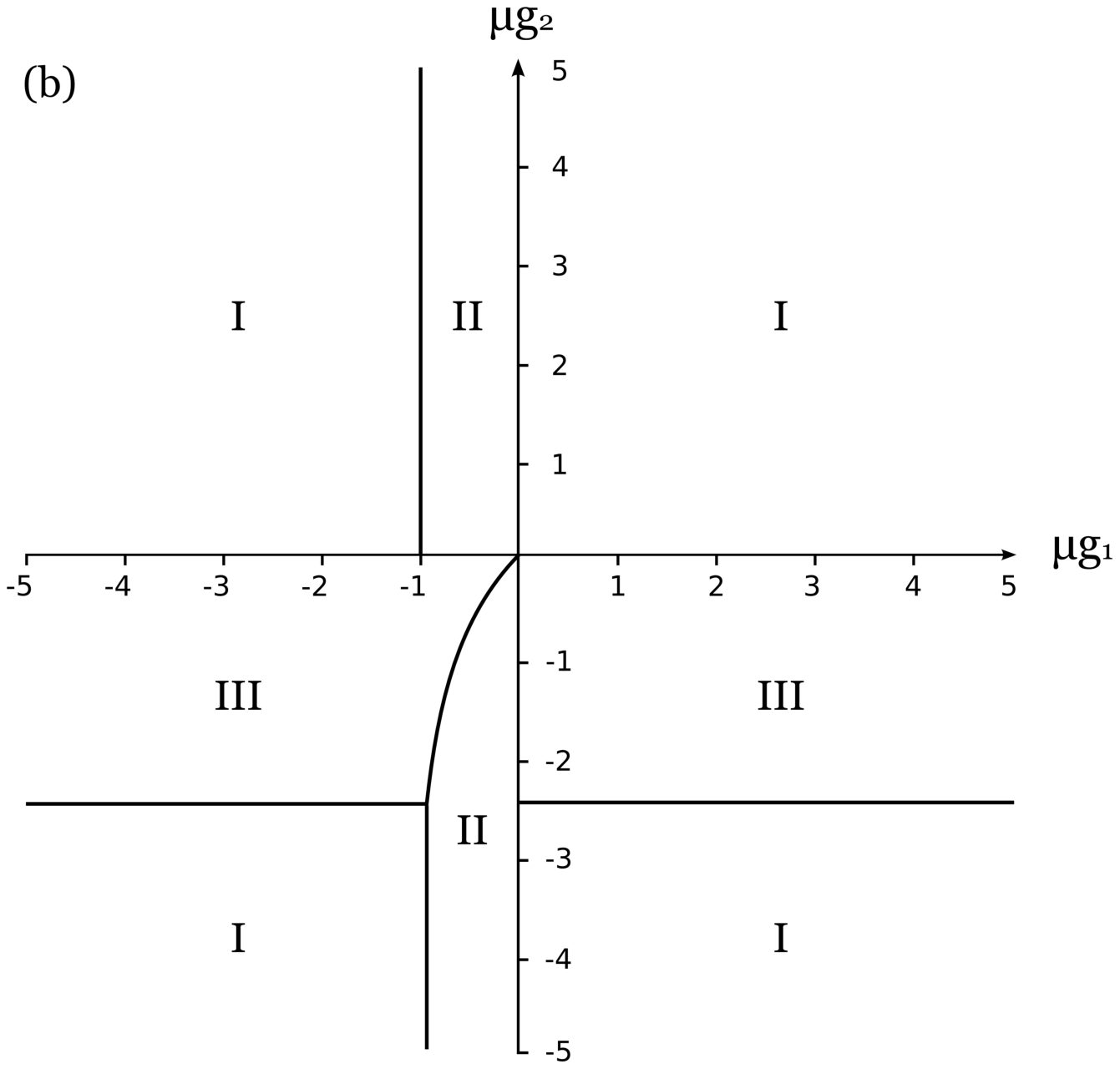}\\
\includegraphics[width=0.45\textwidth]{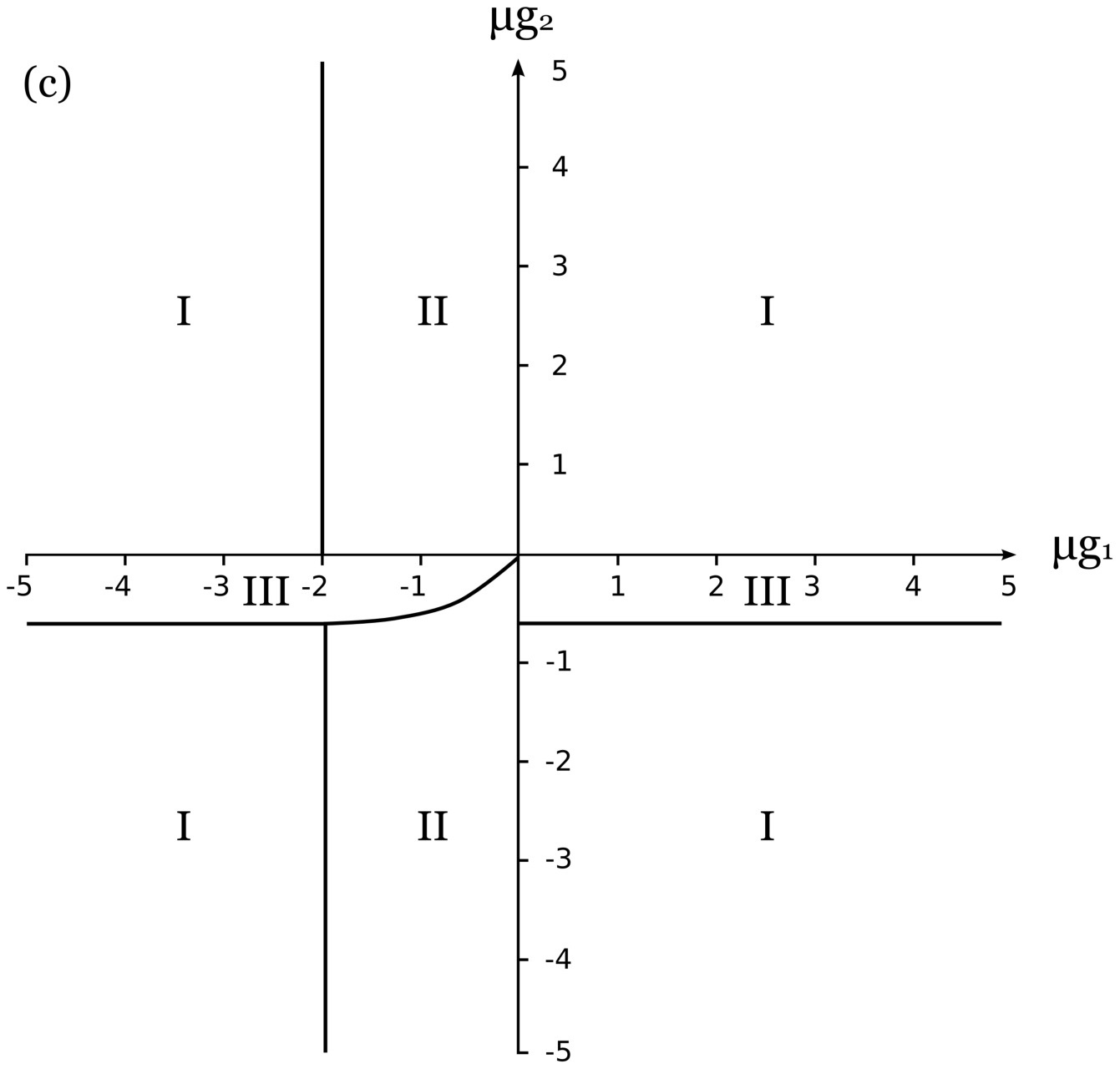}
\hfill
\includegraphics[width=0.45\textwidth]{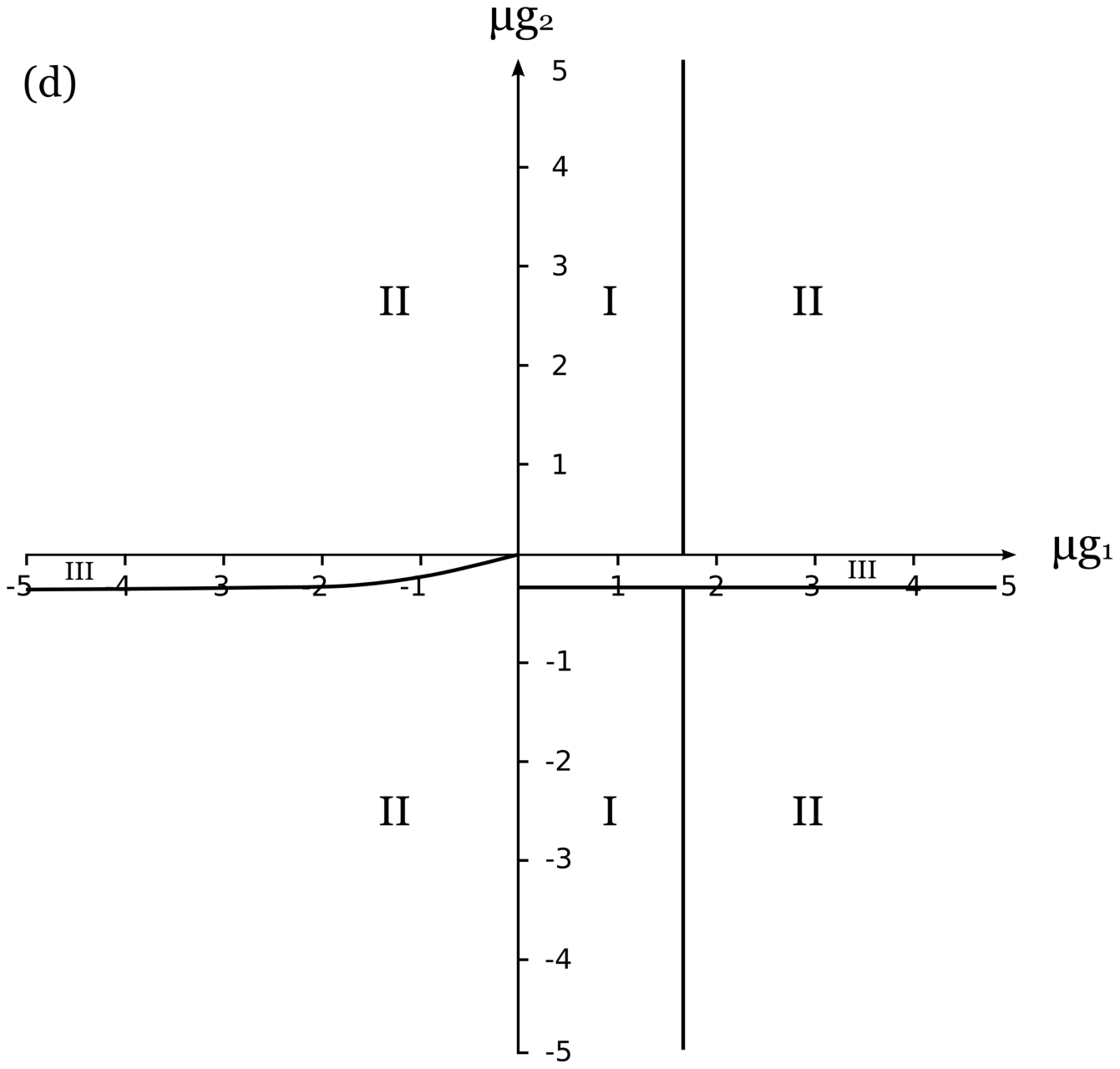}\\
\caption{The $(g_1,g_2)$-phase portrait of the model at arbitrary
fixed values of $\mu$ and for different values of the chiral chemical potential $\mu_5$. (a) The case $\mu_5=0.2\mu$. (b) The case  $\mu_5=0.7\mu$. (c) The case  $\mu_5=1.5\mu$. (d) The case  $\mu_5=2.5\mu$. We use the same designations of the phases as in Fig. 1. }
\end{figure}

\subsubsection{The $(g_1,g_2)$-phase structure at some nontrivial values
of $\mu$ and $\mu_5$}\label{IVC}

Let us now fix the fermion number chemical  potential $\mu\ne 0$ and
consider how the $(g_1,g_2)$-phase portrait of the model is evolved
vs the chiral chemical potential $\mu_5$. This phase portrait at
$\mu_5=0$ is presented in Fig. 5. In Figs. 7 we have drawn several
$(g_1,g_2)$-phase portraits at different relations between $\mu_5$
and $\mu$, (a) $\mu_5=0.2\mu$, (b) $\mu_5=0.7\mu$, (c)
$\mu_5=1.5\mu$, (d) $\mu_5=2.5\mu$. Analyzing Figs. 5 and 7, we see
the following phase evolution of the system vs $\mu_5$ at fixed
$\mu\ne 0$. At $\mu_5=0$ the SC phase III fills almost the whole
$(\mu g_1,\mu g_2)$ plane (see Fig. 5). It means that at arbitrary
fixed $g_1\ne 0$ and $g_2\ne 0$ we have superconductivity in the
system at sufficiently high values of $\mu$ (see Section \ref{IVB}).
If $\mu_5$ begins to grow, then for a lot of $(g_1,g_2)$ points
there might occur a restoration of the initial symmetry (the
symmetrical phase I arises) in the model. However, the further
growth of $\mu_5$ leads to the final appearance of the CSB phase. It
is the general property of the chiral chemical potential $\mu_5$,
which promotes the chiral symmetry breaking in the system.

Since the above mentioned constraints (a),..,(d) at which  Figs. 7
were drawn are not invariant under the duality transformation ${\cal
D}$ (\ref{16}), we would like to note that the phase portraits of
Fig. 7 are not selfdual. In this case the possible duality
transformation (see the previous subsubsection \ref{IVB}) of Figs. 7
might result in the set of four phase portraits, which are a good
illustration of the evolution of the model phase structure vs the
particle number chemical potential $\mu$ at arbitrary fixed
$\mu_5\ne 0$. It is straightforwardly clear from the dual mapping of
Figs. 7 that at arbitrary fixed values of $\mu_5>0$ and $g_1\ne 0$,
$g_2\ne 0$ we have in the system the appearance of superconductivity
at sufficiently high $\mu$. This property is dually conjugated to
the appearing of CSB phase at sufficiently high values of $\mu_5$,
which is a consequence of the phase portraits Fig. 7 (see the above
consideration).

\subsubsection{The $(\mu,\mu_5)$-phase structure at some nontrivial
relations between $g_1$ and $g_2$}

We have already considered the $(\mu,\mu_5)$-phase  structure of the
model in the case of dually symmetrical constraints on the coupling
constants, i.e. when $g_1=g_2$ (see Figs. 3, 4). In the present
Section, special  attention is paid to a more general case when
$g_1\ne g_2$. Here we construct $(\mu,\mu_5)$-phase portraits of the
model for qualitatively different representative relations between
$g_1$ and $g_2$. Namely, we fix in each of the regions of Fig. 1
some representative $(g_1,g_2)$ point and draw a corresponding
$(\mu,\mu_5)$-phase diagram. (Note that due to a duality symmetry of
the model, it is sufficient to reduce the number of the
representative  $(g_1,g_2)$ points to three.)

In Fig. 8 the $(\mu,\mu_5)$-phase portrait corresponds to arbitrary
fixed $g_1>0$ and $g_2=0.2 g_1$ (it is clear that at $\mu=\mu_5=0$
the point $(g_1,g_2=0.2 g_1)$ lies in the symmetrical region I of
Fig. 1). On the phase boundaries of this figure there are phase
transitions of the second order. In Fig. 9 the $(\mu,\mu_5)$-phase
portrait corresponds to arbitrary fixed $g_1<0$ and $g_2=-2 g_1$ (in
this case at $\mu=\mu_5=0$ the point $(g_1,g_2=-2 g_1)$ lies in the
CSB region II of Fig. 1). Finally, in Fig. 10 the
$(\mu,\mu_5)$-phase portrait corresponds to arbitrary fixed $g_1<0$
and $g_2=0.5 g_1$ (in this case at $\mu=\mu_5=0$ the point
$(g_1,g_2=0.5 g_1)$ lies in the SC region III of Fig. 1). To obtain
the $(\mu,\mu_5)$-phase portraits of the model when the point
$(g_1,g_2)$ is fixed in the two remaining areas of Fig. 1, it is
enough to perform the dual mapping of the Figs 9 and 10.

In Figs. 9, 10 all phase boundaries are the curves of the first order
phase transitions, so the point T on these figures is a triple point,
where three phases coexist. The main conclusion from Figs 8-10 is that
the growth of $\mu_5$ (the growth of $\mu$) induces the chiral
symmetry breaking (induces the superconductivity).

\begin{figure}
\includegraphics[width=0.45\textwidth]{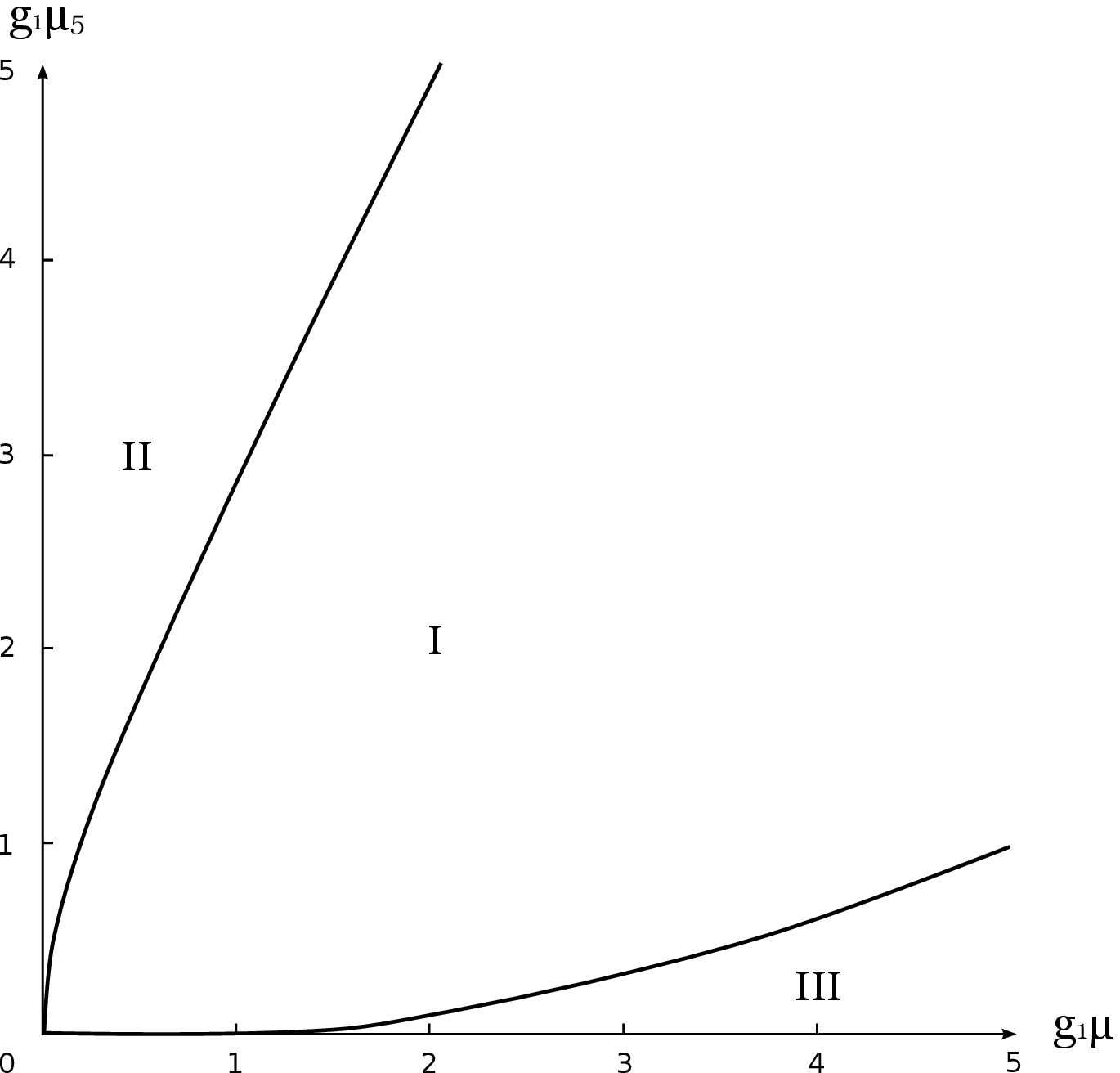}
\hfill
\includegraphics[width=0.45\textwidth]{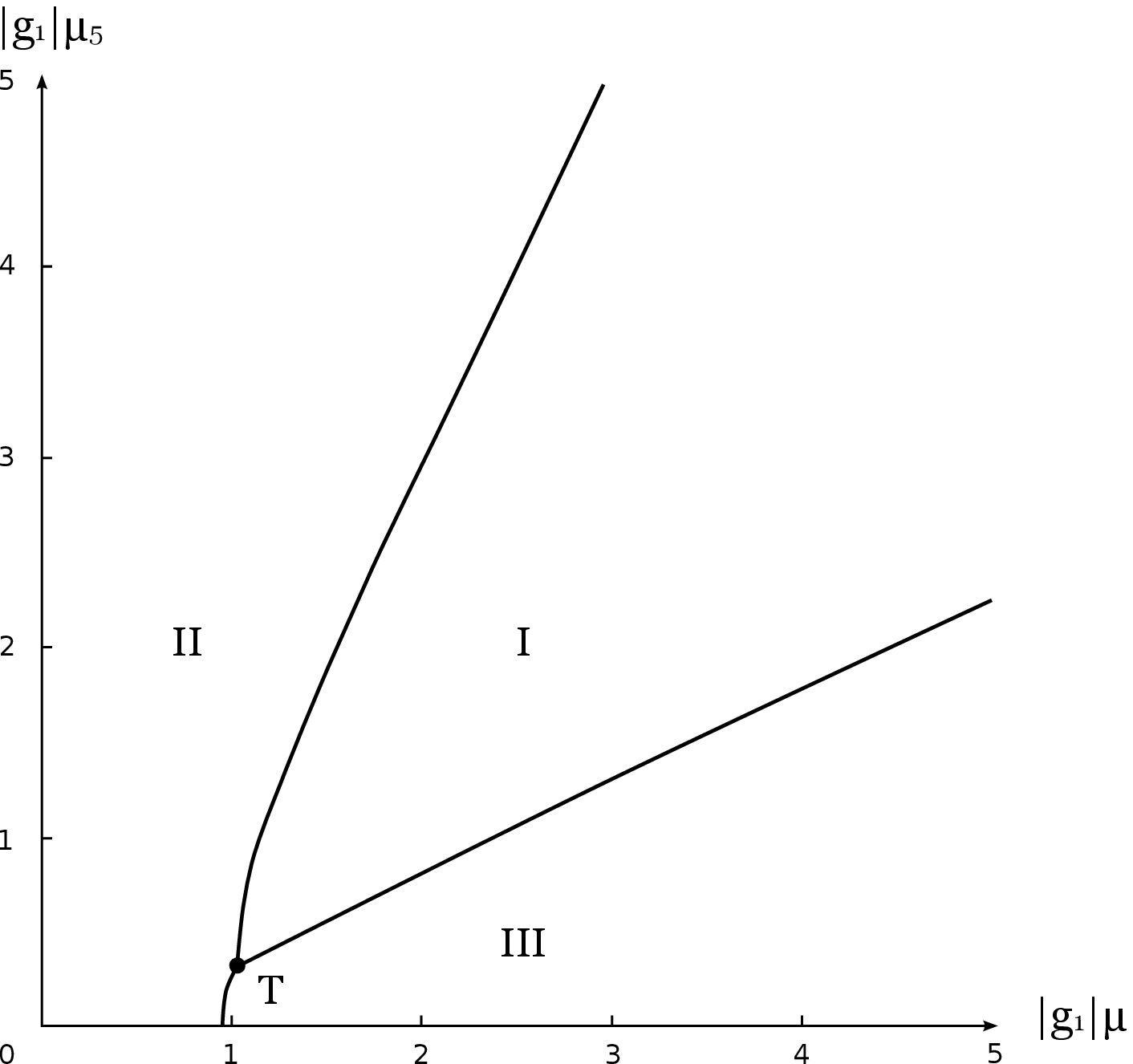}
\\
\parbox[t]{0.45\textwidth}{
\caption{The $(\mu,\mu_5)$-phase portrait of the model at arbitrary fixed $g_1>0$ and $g_2=0.2 g_1$. The notations I, II and III are the same as in Fig. 1.} } \hfill
\parbox[t]{0.45\textwidth}{
\caption{The $(\mu,\mu_5)$-phase portrait of the model at arbitrary fixed
$g_1<0$ and $g_2=-2g_1$. The notations I,
II and III are the same as in Fig. 1. The letter T denotes a triple point, $\mu_T\approx 1/|g_1|$, $\mu_{5T}\approx 0.3/|g_1|$.}}
\end{figure}
\begin{figure}
\includegraphics[width=0.45\textwidth]{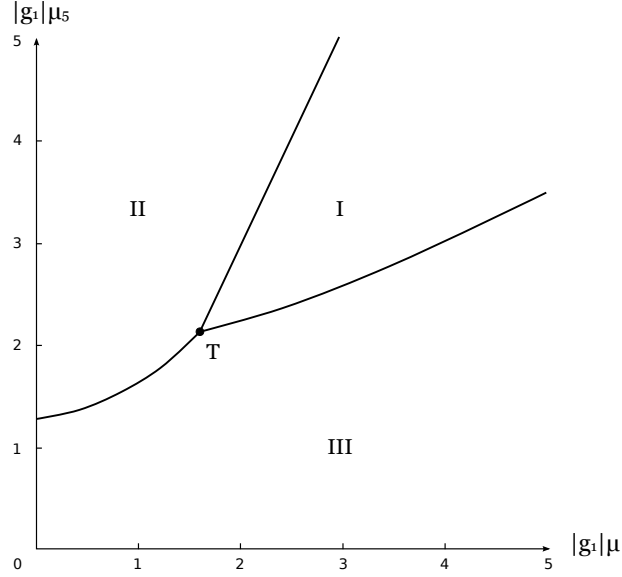}
\\
\caption{The $(\mu,\mu_5)$-phase portrait of the model at arbitrary fixed $g_1<0$ and $g_2=0.5 g_1$. The notations I, II and III are the same as in Fig. 1. The letter T denotes a triple point, $\mu_T\approx 1.6/|g_1|$, $\mu_{5T}\approx 2.2/|g_1|$.}
\end{figure}

\section{Summary and conclusions}

In this paper, the duality correspondence phenomenon between CSB and SC, which was found earlier in (1+1)-dimensional 4F theories \cite{oj,vas,thies1,ekkz},
is demonstrated to take place also in the framework of the
(2+1)-dimensional 4F  model (1). (An alternative 4F model with other
dually conjugated chiral and superconducting channels of interaction
is considered in Appendix \ref{ApD}). It contains both
fermion-antifermion (with $G_1$ a bare coupling constant) and difermion (with $G_2$ a bare coupling constant) interaction channels
and two kinds of chemical potentials, $\mu,\mu_5$. The duality
property of the model means that its thermodynamic potential
$\Omega(M,\Delta)$ (\ref{11}) is invariant under the duality
transformation (\ref{16}). As a result, we have established the
duality correspondence between CSB and SC phases of the
model (1) (as well as of the alternative model of Appendix
\ref{ApD}). Let us suppose, e.g., that for some fixed set
$(g_1,g_2,\mu,\mu_5)$ of external parameters (the coupling constants
$g_1$ and $g_2$ are connected with bare couplings $G_{1,2}$ by the
relation (\ref{13})) the chiral symmetry breaking phase is realized
in the model. Then for rearranged values of external parameters,
$g_1\leftrightarrow g_2$, $\mu\leftrightarrow\mu_5$, (or for dually
conjugated set $(g_2,g_1,\mu_5,\mu)$) we have the so-called dually
conjugated superconducting phase (and vice versa). Moreover, it must
be emphasized that the chiral condensate of the CSB phase, realized for the set $(g_1,g_2,\mu,\mu_5)$, is equal to the superconducting  condensate of the dually conjugated SC phase, corresponding to the set of external
parameters with $g_1\leftrightarrow g_2$, $\mu\leftrightarrow\mu_5$
(and vice versa). In this way, it is sufficient to have the
information about the ground state of the initial phase, which is
realized for the set $(g_1,g_2,\mu,\mu_5)$, in order to determine the properties of the ground state of the dually conjugated phase, corresponding to the rearranged external parameter set $(g_2,g_1,\mu_5,\mu)$.

In order to study the role and influence of the duality property on
the phase structure of the model, for comparison and illustrations, we have demonstrated a variety of phase portraits in the ($\mu,\mu_5$)- and $(g_1,g_2)$ planes. In particular, if the constraint, under which we obtain a phase portrait, is dually invariant, i.e. at $\mu=\mu_5$ or at $g_1=g_2$, then we have a selfdual phase diagram (see, e.g., Figs 1--4). The selfdual phase portrait is mapped into itself under the duality
transformation (it is introduced in Sec. \ref{IVB}) , and its most
characteristic feature is the mirror symmetrical arrangement of the
CSB and SC phases with respect to the line $\mu=\mu_5$ (or
$g_1=g_2$) of the phase diagram.

We have also presented a series of nonselfdual phase portraits,
which do not transform into themselves under the duality mapping
(see Figs. 5--10). The results obtained lead to the conclusion that
the growth of the chiral chemical potential $\mu_5$ promotes the
chiral symmetry breaking,
\footnote{The properties of the simplest (2+1)-dimensional GN and (3+1)-dimensional NJL models under the
influence of $\mu_5$ were studied, correspondingly, in the recent papers \cite{Cao} and \cite{Braguta}. It was shown there that the chiral
chemical potential $\mu_5$ plays a role of the catalyst of dynamical chiral symmetry breaking. }
whereas particle number chemical potential
$\mu$ induces superconductivity in the system (see also in
\cite{kzz}).

We hope that our investigations may shed some new light on physical
effects in planar systems like high-temperature superconductors or
graphene.

\appendix
\section{Algebra of the $\gamma$ matrices in the case of SO(2,1) group}
\label{ApA}

The two-dimensional irreducible representation of the (2+1)-dimensional
Lorentz group SO(2,1) is realized by the following $2\times 2$
$\tilde\gamma$-matrices:
\begin{eqnarray}
\tilde\gamma^0=\sigma_3=
\left (\begin{array}{cc}
1 & 0\\
0 &-1
\end{array}\right ),\,\,
\tilde\gamma^1=i\sigma_1=
\left (\begin{array}{cc}
0 & i\\
i &0
\end{array}\right ),\,\,
\tilde\gamma^2=i\sigma_2=
\left (\begin{array}{cc}
0 & 1\\
-1 &0
\end{array}\right ),
\label{A1}
\end{eqnarray}
acting on two-component Dirac spinors. They have the properties:
\begin{eqnarray}
Tr(\tilde\gamma^{\mu}\tilde\gamma^{\nu})=2g^{\mu\nu};~~
[\tilde\gamma^{\mu},\tilde\gamma^{\nu}]=-2i\varepsilon^{\mu\nu\alpha}
\tilde\gamma_{\alpha};~
~\tilde\gamma^{\mu}\tilde\gamma^{\nu}=-i\varepsilon^{\mu\nu\alpha}
\tilde\gamma_{\alpha}+g^{\mu\nu},
\label{A2}
\end{eqnarray}
where $g^{\mu\nu}=g_{\mu\nu}=diag(1,-1,-1),
~\tilde\gamma_{\alpha}=g_{\alpha\beta}\tilde\gamma^{\beta},~
\varepsilon^{012}=1$.
There is also the relation:
\begin{eqnarray}
Tr(\tilde\gamma^{\mu}\tilde\gamma^{\nu}\tilde\gamma^{\alpha})=
-2i\varepsilon^{\mu\nu\alpha}.
\label{A3}
\end{eqnarray}
Note that the definition of chiral symmetry is slightly unusual in
(2+1)-dimensions (spin is here a pseudoscalar rather than a (axial)
vector). The formal reason is simply that there exists no other $2\times 2$ matrix anticommuting with the Dirac matrices $\tilde\gamma^{\nu}$
which would allow the introduction of a $\gamma^5$-matrix in the
irreducible representation. The important concept of 'chiral'
symmetries  and their breakdown by mass terms can nevertheless be
realized also in the framework of (2+1)-dimensional quantum field
theories by considering a four-component reducible representation
for Dirac fields. In this case the Dirac spinors $\psi$ have the
following form:
\begin{eqnarray}
\psi(x)=
\left (\begin{array}{cc}
\tilde\psi_{1}(x)\\
\tilde\psi_{2}(x)
\end{array}\right ),
\label{A4}
\end{eqnarray}
with $\tilde\psi_1,\tilde\psi_2$ being two-component spinors.
In the reducible four-dimensional spinor representation one deals
with 4$\times$4 $\gamma$-matrices:
$\gamma^\mu=diag(\tilde\gamma^\mu,-\tilde\gamma^\mu)$, where
$\tilde\gamma^\mu$ are given in (\ref{A1}). One can easily show, that
($\mu,\nu=0,1,2$):
\begin{eqnarray}
&&Tr(\gamma^\mu\gamma^\nu)=4g^{\mu\nu};~~
\gamma^\mu\gamma^\nu=\sigma^{\mu\nu}+g^{\mu\nu};~~\nonumber\\
&&\sigma^{\mu\nu}=\frac{1}{2}[\gamma^\mu,\gamma^\nu]
=diag(-i\varepsilon^{\mu\nu\alpha}\tilde\gamma_\alpha,
-i\varepsilon^{\mu\nu\alpha}\tilde\gamma_\alpha).
\label{A5}
\end{eqnarray}
In addition to the  Dirac matrices $\gamma^\mu~~(\mu=0,1,2)$ there
exist two other matrices, $\gamma^3$ and $\gamma^5$, which anticommute
with all $\gamma^\mu~~(\mu=0,1,2)$ and with themselves
\begin{eqnarray}
\gamma^3=i\left (\begin{array}{cc}
0~,&- I\\
I~,& 0
\end{array}\right ),\,\,
\gamma^5=\left (\begin{array}{cc}
0~,& I\\
I~,& 0
\end{array}\right ),\,
\label{A6}
\end{eqnarray}
with  $I$ being the unit $2\times 2$ matrix. It is obvious that $\gamma^3= \gamma^0\gamma^1\gamma^2\gamma^5$ and $\gamma^5= -\gamma^0\gamma^1\gamma^2\gamma^3$.

\section{Alternative 4F (2+1)-dimensional model with dual symmetry
between other CSB and SC channels}
\label{ApD}

Each of the matrices $\gamma^3$ and $\gamma^5$  (see Appendix
\ref{ApA}) can be selected as a generator for the corresponding
$U_{\gamma^3}(1)$ and $U_{\gamma^5}(1)$ chiral group of spinor field
transformations. For example, the Lagrangian (1) is invariant with
respect to $U_{\gamma^5}(1)$ such that
$\psi_k(x)\to\exp(i\alpha\gamma^5)\psi_k(x)$. Alternatively, it is
possible to construct a 4F model with fermion-antifermion and
superconducting channels, symmetric under $U_{\gamma^3}(1)$
continuous chiral transformations,
$\psi_k(x)\to\exp(i\alpha\gamma^3)\psi_k(x)$ ($k=1,..,N$). Its
Lagrangian, e.g., reads
\begin{eqnarray}
 L_{\gamma^3}\equiv L_{\gamma^3}(G_1,G_2;\mu,\mu_3)=\sum_{k=1}^{N}\bar \psi_k\Big [\gamma^\nu i\partial_\nu
+\mu\gamma^0+\mu_3\gamma^0\gamma^3\Big ]\psi_k&+& \frac {G_1}N\left (\widetilde {4F}\right )_{ch}+\frac {G_2}N\left
(\widetilde {4F}\right )_{sc},\label{A7}
\end{eqnarray}
where the four-fermion structures  $\left (\widetilde {4F}\right
)_{ch}$ and $\left (\widetilde {4F}\right )_{sc}$ are used,
\begin{eqnarray}
\left (\widetilde {4F}\right )_{ch}=\left
(\sum_{k=1}^{N}\bar \psi_k\psi_k\right )^2+\left (\sum_{k=1}^{N}\bar \psi_k
i\gamma^3\psi_k\right )^2,~~\left(\widetilde {4F}\right)_{sc}=\left
(\sum_{k=1}^{N} \psi_k^T \widetilde C\psi_k\right )\left (\sum_{j=1}^{N}\bar
\psi_j \widetilde C\bar\psi_j^T\right ). \label{A71}
\end{eqnarray}
Here $\widetilde C=iC\gamma^3\gamma^5$ and  $\mu$ is the usual
particle number chemical potential (as in (1)). Since this
Lagrangian is invariant under $U_{\gamma^3}(1)$, there exist a
corresponding conserved density of chiral charge
$n_3=\sum_{k=1}^{N}\bar \psi_k\gamma^0\gamma^3\psi_k$ as well as its
thermodynamically conjugate quantity, the chiral (or axial) chemical
potential $\mu_3$. Note that in the framework of the model
(\ref{A7}) there is also the duality between chiral and
superconducting channels of interaction, i.e. between $\left
(\widetilde {4F}\right )_{ch}$ and $\left(\widetilde
{4F}\right)_{sc}$ four-fermion structures (\ref{A71}). Indeed,
performing in (\ref{A7})-(\ref{A71}) the modified Pauli-Gursey
transformation of spinor fields,
\begin{eqnarray}
\widetilde{PG}:~~\psi_k (x)
\longrightarrow \frac 12 (1-\gamma^3)\psi_k (x)+\frac 12
(1+\gamma^3)\widetilde C\bar\psi^T_k(x), \label{D1}
\end{eqnarray}
we see that
\begin{eqnarray}
\left(\widetilde {4F}\right)_{ch}\stackrel{
\widetilde{PG}}{\longleftrightarrow} \left(\widetilde {4F}\right)_{sc},~~~  L_{\gamma^3}(G_1,G_2;\mu,\mu_3) \stackrel{\widetilde{PG}}{\longleftrightarrow} L_{\gamma^3}(G_2,G_1;-\mu_3,-\mu).\label{D2}
\end{eqnarray}
The relations (\ref{D2}) are the basis for  the duality invariance
of the thermodynamic potential of the model (\ref{A7}).

To verify this, one can use the way of Sec. II. Indeed, the
semi-bosonized version of Lagrangian (\ref{A7}) that contains
only quadratic powers of fermionic fields and auxiliary  bosonic fields $\sigma (x)$, $\tilde\pi(x)$, $\tilde\Delta(x)$ and $\tilde\Delta^{*}(x)$ has the following form
\begin{eqnarray}
{\cal L}_{\gamma^3}\ds &=&\bar\psi_k\Big [\gamma^\nu i\partial_\nu +\mu\gamma^0
+\mu_3\gamma^0\gamma^3-\sigma -i\gamma^3\tilde\pi\Big ]\psi_k
 -\frac{N(\sigma^2+\tilde\pi^2)}{4G_1} \nonumber\\
&&~~~~~~~~~~~~~~~~~-\frac N{4G_2}\tilde\Delta^{*}\tilde\Delta-
 \frac{\tilde\Delta^{*}}{2}[\psi_k^T\widetilde C\psi_k]
-\frac{\tilde\Delta}{2}[\bar\psi_k \widetilde C\bar\psi_k^T] \label{A8}
\end{eqnarray}
(in (\ref{A8}) and below the summation over $k=1,...,N$ is implied). On Euler-Lagrange equations for bosonic fields,
\begin{eqnarray}
\sigma (x)=-2\frac {G_1}N(\bar\psi_k\psi_k),&~~&\tilde\pi(x)=-2\frac {G_1}N(\bar\psi_k i\gamma^3\psi_k),~~ \nonumber\\
\tilde\Delta(x)=-2\frac
{G_2}N(\psi_k^T\widetilde C\psi_k),&~~& \tilde\Delta^{*}(x)=-2\frac
{G_2}N(\bar\psi_k\widetilde C\bar\psi_k^T), \label{A9}
\end{eqnarray}
the semi-bosonized Lagrangian (\ref{A8})  is equivalent to the 4F
Lagrangian (\ref{A7}). Note also that the bosonic fields
$\tilde\Delta(x)$ and $\tilde\Delta^*(x)$ are
pseudoscalars, whereas $\sigma (x)$ and $\tilde\pi (x)$ are scalars
with respect to parity transformation (\ref{03}).

Then, the thermodynamic potential $\Omega_{\gamma^3} (M,\Delta)$, which describes the ground state of the model (\ref{A7})-(\ref{A8}), is analogously defined by the following relation (for convenience, we omit here and below the tilde in fields)
\begin{eqnarray}
\int d^3x\Omega_{\gamma^3} (M,\Delta)\,\,&=&\,\,\int d^3x\left
(\frac{M^2}{4G_1}+\frac{\Delta^2}{4G_2}\right )+\frac{i}{N}\ln\left
( \int\prod_{l=1}^{N}[d\bar\psi_l][d\psi_l]\exp\Big (i\int d^3 x\Big
[\bar\psi_k D_{\gamma^3}\psi_k\right.\nonumber\\&& \left.-
\frac{\Delta}{2}(\psi_k^T\widetilde C\psi_k)
-\frac{\Delta}{2}(\bar\psi_k\widetilde C\bar\psi_k^T)\Big ]
\Big )\right ), \label{90}
\end{eqnarray}
where $D_{\gamma^3}=\gamma^\rho  i\partial_\rho+\mu\gamma^0
+\mu_3\gamma^0\gamma^3-M$. In writing down this expression, we used
the same arguments that were used in the derivation of the
thermodynamic potential in the framework of the model (1) (see the
derivation of (\ref{9}) in Section II). Path integration in the
expression (\ref{90}) is evaluated in Appendix \ref{ApB} so we have
for the TDP (\ref{90}) the following expression
\begin{eqnarray}
\Omega_{\gamma^3} (M,\Delta)&=&
\frac{M^2}{4G_1}+\frac{\Delta^2}{4G_2}
+\frac{i}{2}\int\frac{d^3p}{(2\pi)^3}\ln\Big
[\tilde\lambda_1(p)\tilde\lambda_2(p)\tilde\lambda_3(p)\tilde\lambda_4(p)\Big ], \label{91}
\end{eqnarray}
where $\tilde\lambda_{1,...,4}(p)$ are  presented in (\ref{B14}).
Comparing this thermodynamic potential with the TDP $\Omega
(M,\Delta)$ (\ref{11}) of the model (1), we see that
\begin{eqnarray}
\Omega_{\gamma^3} (M,\Delta)&=&\Omega (M,\Delta)\Big |_{\mu_5\to\mu_3}.
 \label{92}
\end{eqnarray}
As a result, it is clear that the TDP  $\Omega_{\gamma^3}
(M,\Delta)$ of the 4F model (\ref{A7}) is invariant under the
following dual transformation
\begin{eqnarray}
G_1\longleftrightarrow G_2,~~M\longleftrightarrow \Delta,~~\mu\longleftrightarrow\mu_3.
 \label{160}
\end{eqnarray}
Furthermore, to find phase portraits of  the model (\ref{A7}), it is
sufficient to perform in all Figs 1,..,10 the replacement
$\mu_5\to\mu_3$.

\section{The path integration over anticommuting fields}
\label{ApB}

Let us calculate the following path integral over anticommuting
four-component Dirac spinor fields $\psi(x)$, $\bar\psi(x)$:
\begin{eqnarray}
I_\Gamma=\int[d\bar\psi][d\psi]
\exp\Big (i\int d^3 x\Big
[\bar\psi  D_\Gamma\psi-\frac{\Delta}{2}(\psi^TC\Gamma\psi) -\frac{\Delta}{2}(\bar\psi C\Gamma\bar\psi^T)\Big ] \Big ), \label{B1}
\end{eqnarray}
where $C$ is the charge conjugation  matrix, $C=\gamma^2$. Since in
the 4$\times$4 spinor space the product matrix $C\Gamma$ should be
an antisymmetric one, it is clear that the $\Gamma$ matrix in
(\ref{B1}) might be, e.g., the unit matrix ${\bf 1_s}$ in the spinor
space or $\Gamma=i\gamma^3\gamma^5$ (for notations see Appendix
\ref{ApA}), etc. Note in addition that at $\Gamma={\bf 1_s}$ and
$D_\Gamma=D$ (see in (\ref{9})) the integral $I_\Gamma$ is equal to
the argument of the $\ln$-function in the formula (\ref{9}) in the
particular case $N=1$, whereas at $\Gamma=i\gamma^3\gamma^5$ and
$D_\Gamma=D_{\gamma^3}$ (see (\ref{90})) the integral $I_\Gamma$ is
equal to the argument of the $\ln$-function in the formula
(\ref{90}) also in the particular case $N=1$. Recall, there are
general Gaussian path integrals \cite{vasiliev}:
\begin{eqnarray}
\int[d\psi]\exp\Big (i\int d^3 x\Big
[-\frac{1}{2}\psi^T A \psi +\eta^T\psi
\Big ]\Big )&=&\left(\det(A)\right )^{1/2}\exp\Big (-\frac{i}{2}\int d^3 x\Big [\eta^T A^{-1}\eta\Big ]\Big ),  \label{B2}\\
\int[d\bar\psi]\exp\Big (i\int d^3 x\Big
[-\frac{1}{2}\bar\psi A \bar\psi^T +\bar\eta\bar\psi^T\Big ] \Big )&
=&\left (\det(A)\right )^{1/2}\exp\Big (-\frac{i}{2}\int d^3 x\Big
[\bar\eta A^{-1}\bar\eta^T\Big ]\Big ), \label{B3}
\end{eqnarray}
where $A$ is an antisymmetric operator  in coordinate and spinor
spaces, and $\eta(x)$, $\bar\eta (x)$ are anticommuting spinor
sources which also anticommute with $\psi$ and $\bar\psi$. First,
let us integrate in (\ref{B1}) over $\psi$-fields with the help of
the relation (\ref{B2}) supposing there that $A=\Delta C\Gamma$,
$\bar\psi D_\Gamma=\eta^T$, i.e. $\eta=D_\Gamma^T\bar\psi^T$. Then
\begin{eqnarray}
I_\Gamma=\left (\det(\Delta C\Gamma)\right )^{1/2}\int[d\bar\psi]\exp\Big (-\frac{i}{2}\int d^3 x
\bar\psi \big [\Delta C\Gamma+D_\Gamma(\Delta C\Gamma)^{-1}D_\Gamma^T\big ]\bar\psi^T\Big ). \label{B4}
\end{eqnarray}
Second, the integration over  $\bar\psi$-fields in (\ref{B4}) can be
easily performed with the help of the formula (\ref{B3}), where one
should put $A=\Delta C\Gamma+D_\Gamma(\Delta
C\Gamma)^{-1}D_\Gamma^T$ and $\bar\eta=0$. As a result, we have
\begin{eqnarray}
I_\Gamma=\left (\det(\Delta C\Gamma)\right )^{1/2}\left (\det[\Delta C\Gamma+D_\Gamma(\Delta C\Gamma)^{-1}D_\Gamma^T]\right )^{1/2}=\left (\det [\Delta^2(C\Gamma)^2+D_\Gamma (C\Gamma)^{-1} D_\Gamma^T C\Gamma]\right )^{1/2}. \label{B5}
\end{eqnarray}
We evaluate the expression (\ref{B5}) in  two cases, (i) when
$\Gamma={\bf 1_s}$ and $D_\Gamma=D$, where $D$ is presented in
(\ref{9}), and (ii) when $\Gamma=i\gamma^3\gamma^5$ and
$D_\Gamma=D_{\gamma^3}$, where $D_{\gamma^3}$ is given in
(\ref{90}).

{\bf (i) The case $\Gamma ={\bf 1_s}$ and $D_\Gamma=D=\gamma^\rho i\partial_\rho +\mu\gamma^0+\mu_5\gamma^0\gamma^5-M$}.
Taking into account the relations $(\partial_\nu)^T=-\partial_\nu$
and $C^{-1} (\gamma^\nu)^T C =-\gamma^\nu$ ($\nu=0,1,2$), we obtain
from (\ref{B5})
\begin{eqnarray}
I_{{\bf 1_s}}=\left (\det [-\Delta^2+D_+D_-]\right )^{1/2}\equiv\left (\det B\right )^{1/2}, \label{B7}
\end{eqnarray}
where $D_\pm=\gamma^\nu i\partial_\nu-M\pm (\mu\gamma^0+ \mu_5\gamma^0 \gamma^5)$. Using the general relation $\det B =\exp ({\rm Tr}\ln B)$, we get from (\ref{B7}):
\begin{eqnarray}
\ln I_{{\bf 1_s}}=\frac 12 {\rm Tr}\ln\left (B\right
)=\frac 12\sum_{i=1}^{4}\int\frac{d^3p}{(2\pi)^3} \ln(\lambda_i(p))\int
d^3x. \label{B8}
\end{eqnarray}
(A more detailed consideration of operator traces is presented in
Appendix A of the paper \cite{Ebert:2009ty}.) In this formula the
symbol Tr means the trace of an operator both in the coordinate and
internal spaces. Moreover, $\lambda_i(p)$ ($i=1,...,4$) in (\ref{B8})
are four eigenvalues of the 4$\times$4 Fourier
transformation matrix $\bar B(p)$ of the operator $B$ from (\ref{B7}). Namely,
\begin{eqnarray}
\lambda_{1,...,4}(p)&=&M^2-|\vec p|^2+p_0^2-\mu^2+\mu_5^2-\Delta^2 \pm
2\sqrt{M^2(p_0^2-|\vec p|^2)+(|\mu | |\vec p|\pm |\mu_5|p_0)^2},
\label{B12}
\end{eqnarray}
where $|\vec p|=\sqrt{p_1^2+p_2^2}$. It is clear from these relations that $\lambda_{1,...,4}(p)$ are even functions vs $\mu$ and/or $\mu_5$. So, it is enough to take into account only nonnegative values of the chemical potentials $\mu$ and $\mu_5$.

{\bf (ii) The case $\Gamma =i\gamma^3\gamma^5$  and
$D_\Gamma=D_{\gamma^3}=\gamma^\rho i\partial_\rho+\mu\gamma^0
+\mu_3\gamma^0\gamma^3-M$}. In this case
$C^{-1}D_\Gamma^TC=\gamma^\rho i\partial_\rho-\mu\gamma^0
+\mu_3\gamma^0\gamma^3-M$. Then it follows from (\ref{B5}) that
\begin{eqnarray}
I_\Gamma=\left (\det [-\Delta^2+\tilde D_+\tilde D_-]\right )^{1/2}\equiv\left (\det {\cal B}\right )^{1/2}, \label{B10}
\end{eqnarray}
where $\tilde D_\pm=\gamma^\nu i\partial_\nu-M\pm ( \mu\gamma^0+\mu_3\gamma^0\gamma^3)$. Then, similar to (\ref{B8}), it is easy to find that in the case under consideration
\begin{eqnarray}
\ln I_{\Gamma}=\frac 12 {\rm Tr}\ln\left ({\cal B}\right
)=\frac 12\sum_{i=1}^{4}\int\frac{d^3p}{(2\pi)^3} \ln(\tilde \lambda_i(p))\int d^3x, \label{B13}
\end{eqnarray}
where  $\tilde \lambda_i(p)$ ($i=1,...,4$) are  four eigenvalues of
the 4$\times$4 Fourier transformation matrix $\bar {\cal B}(p)$ of
the operator ${\cal B}$ from (\ref{B10}). It turns out that the
eigenvalues $\tilde \lambda_i(p)$ are connected with the eigenvalues
$\lambda_i(p)$ (\ref{B12}) by the simple relations
\begin{eqnarray}
\tilde \lambda_i(p)=\lambda_i(p)\Big |_{\mu_5\to\mu_3}. \label{B14}
\end{eqnarray}

\section{Evaluation of the function $F_1(M)$ (\ref{31}).}
\label{ApC}

To calculate the improper convergent integral in (\ref{31}) we first
use there a polar coordinate system, i.e. $\int d^2p=2\pi\int_0^\infty
pdp$, and then restrict the  $p$-integration region by $\Lambda$
(suppose in addition that $\Lambda >>\mu,\mu_5,M$). As a result, we come to the regularized expression $F_{1\Lambda}(M)$ of the TDP (\ref{31}),
$F_{1\Lambda}(M)=V_1(M)+I_1+I_++I_-$, where
\begin{eqnarray}
I_1&=&-\int_0^\Lambda\frac{pdp}{2\pi} \Big [\sqrt{M^2+(p+\mu_5)^2}+\sqrt{M^2+(p-\mu_5)^2}-2\sqrt{p^2+M^2}\Big
],\label{C9}\\
I_\pm&=&-\int_0^\Lambda\frac{pdp}{2\pi}\left (\mu-\sqrt{M^2+(p\pm\mu_5)^2}\right )\theta\left (\mu-\sqrt{M^2+(p\pm\mu_5)^2}\right ).
\label{C10}
\end{eqnarray}
Due to the presence of the $\theta (x)$-functions  in (\ref{C10})
and sufficiently high values of the cutoff parameter
$\Lambda>>\mu,\mu_5,M$, the quantities $I_\pm$ indeed do not depend
on $\Lambda$. Moreover, it is evident that $I_+$ is a nonzero
quantity only in the case $\mu>\sqrt{M^2+\mu_5^2}$. Hence,
substituting $q=p+\mu_5$ for the $I_+$-integration in (\ref{C10}),
we have
\begin{eqnarray}
I_+&=&-\frac{\theta\left (\mu-\sqrt{M^2+\mu_5^2}\right )}{2\pi}\int_{\mu_5}^{\sqrt{\mu^2-M^2}} (q-\mu_5)\left (\mu-\sqrt{M^2+q^2}\right )dq\label{C12}
\end{eqnarray}
(The upper limit in the integral (\ref{C12})  corresponds to a value
of the momentum $p$, where the $\theta$-function of the integrand
for $I_+$ from (\ref{C10}) is equal to zero, i.e. to a value of $p$
determined by the condition $\mu=\sqrt{M^2+(p+\mu_5)^2}$.)

It is convenient to present the quantity  $I_-$ from (\ref{C10}) as
a sum of two integrals, $I_-=i_1+i_2$, where
\begin{eqnarray}
i_1&=&-\int_0^{\mu_5}\frac{pdp}{2\pi}\left (\mu-\sqrt{M^2+(p-\mu_5)^2}\right )\theta\left (\mu-\sqrt{M^2+(p-\mu_5)^2}\right ),\label{C13}\\i_2&=&-\int^\Lambda_{\mu_5}\frac{pdp}{2\pi}\left (\mu-\sqrt{M^2+(p-\mu_5)^2}\right )\theta\left (\mu-\sqrt{M^2+(p-\mu_5)^2}\right ).
\label{C14}
\end{eqnarray}
It is clear that $i_2$ is a nonzero quantity  only at $\mu>M$. Then,
performing in the integral (\ref{C14}) the substitution $q=p-\mu_5$,
we obtain
\begin{eqnarray}
i_2&=&-\frac{\theta\left (\mu-M\right )}{2\pi}\int_{0}^{\sqrt{\mu^2-M^2}} (q+\mu_5)\left (\mu-\sqrt{M^2+q^2}\right )dq.\label{C15}
\end{eqnarray}
Obviously, we have $i_1\equiv 0$ at $\mu<M$.  To evaluate $i_1$ at
other values of $\mu$, we should, first, substitute $q=\mu_5-p$ in
the integral (\ref{C13}) and then consider two different regions of
the parameter $\mu$, (i)
$\omega_1=\{\mu:M<\mu<\sqrt{M^2+\mu_5^2}\}$, (ii)
$\omega_2=\{\mu:\sqrt{M^2+\mu_5^2}<\mu\}$. As a result, we have
\begin{eqnarray}
i_1&=&-\frac{\theta\left (\mu-M\right )\theta\left (\sqrt{M^2+\mu_5^2}-\mu\right )}{2\pi}\int_{0}^{\sqrt{\mu^2-M^2}} (\mu_5-q)\left (\mu-\sqrt{M^2+q^2}\right )dq\nonumber\\
&&-\frac{\theta\left (\mu-\sqrt{M^2+\mu_5^2}\right )}{2\pi}\int_{0}^{\mu_5} (\mu_5-q)\left (\mu-\sqrt{M^2+q^2}\right )dq.\label{C16}
\end{eqnarray}
Summing the expressions (\ref{C12}), (\ref{C15}) and (\ref{C16}), we have
\begin{eqnarray}
&&I_++I_-=-\frac{\mu_5\theta\left (\mu-M\right )\theta\left (\sqrt{M^2+\mu_5^2}-\mu\right )}{\pi}\int_{0}^{\sqrt{\mu^2-M^2}} \left (\mu-\sqrt{M^2+q^2}\right )dq\nonumber\\
&&-\frac{\theta\left (\mu-\sqrt{M^2+\mu_5^2}\right )}{\pi}\left\{\mu_5\int_{0}^{\mu_5} \left (\mu-\sqrt{M^2+q^2}\right )dq+
\int_{\mu_5}^{\sqrt{\mu^2-M^2}} \left (\mu-\sqrt{M^2+q^2}\right )qdq\right\}.\label{C17}
\end{eqnarray}

To calculate the $I_1$-term of the regularized TDP $F_{1\Lambda}(M)$, it is useful to present $I_1$ as a sum of three integrals, each one corresponds to some square root expression of the integrand in (\ref{C9}). Then, substituting $q=p+\mu_5$ and $q=p-\mu_5$ in two of these integrals, it is possible to present $I_1$ in the form
\begin{eqnarray}
-2\pi I_1&=&\left (\int^0_{\mu_5}+\int_0^\Lambda+\int_\Lambda^{\Lambda+\mu_5}\right )(q-\mu_5)\sqrt{q^2+M^2} dq
-2\int_0^\Lambda q\sqrt{q^2+M^2}dq\nonumber\\
&&~~~~~~~~~~~~+\left (\int^0_{-\mu_5}+\int_0^\Lambda+\int_\Lambda^{\Lambda-\mu_5}\right )(q+\mu_5)\sqrt{q^2+M^2} dq\nonumber\\
&=&2\int_0^{\mu_5}(\mu_5-q)\sqrt{q^2+M^2} dq+\int_\Lambda^{\Lambda+\mu_5}(q-\mu_5)\sqrt{q^2+M^2} dq+\int_\Lambda^{\Lambda-\mu_5}(q+\mu_5)\sqrt{q^2+M^2} dq. \label{C4}
\end{eqnarray}
Since $\Lambda >>\mu,\mu_5,\Delta$, we an use $\sqrt{q^2+\Delta^2}=q+\Delta^2/(2q)+\cdots$ in the last two terms of (\ref{C4}). Then, after integrations one can see that the sum of these two last integrals in (\ref{C4}) is equal to zero in the limit $\Lambda\to \infty$. Hence,
\begin{eqnarray}
\lim_{\Lambda\to\infty}I_1&=&-\frac{1}{\pi}\int_0^{\mu_5}dq(\mu_5-q)\sqrt{q^2+M^2}=-\frac{M^3}{3\pi}+\frac{(\mu_5^2+M^2)\sqrt{\mu_5^2+M^2}}{3\pi}\nonumber\\
&&-\frac{\mu_5^2\sqrt{\mu_5^2+M^2}}{2\pi}-\frac{\mu_5 M^2}{2\pi}\ln\left [\frac{\mu_5+\sqrt{\mu_5^2+M^2}}{M}\right ].\label{C11}
\end{eqnarray}
Finally, performing in  (\ref{C17}) trivial table integrations and using the relation
\begin{eqnarray}
F_1(M)=\lim_{\Lambda\to\infty}F_{1\Lambda}(M)=V_1(M)+\lim_{\Lambda\to\infty}I_1+I_++I_-,\label{C0}
\end{eqnarray}
we obtain from (\ref{C0}), (\ref{C17}) and (\ref{C11}) the expression
(\ref{C18}) (the function $V_1(M)$ is given in the text below (\ref{32})).


\begin{thebibliography}{999}


\bibitem{volkov}
D.~Ebert and M.K.~Volkov,
  Z.\ Phys.\ C {\bf 16}, 205 (1983);
D.~Ebert, H.~Reinhardt and M.K.~Volkov,
  Prog.\ Part.\ Nucl.\ Phys.\  {\bf 33}, 1 (1994).

\bibitem{buballa}
M.~Buballa, Phys.\ Rept.\  {\bf 407}, 205 (2005).

\bibitem{njl}
Y. Nambu and G. Jona-Lasinio, Phys. Rev.{\bf 122}, 345 (1961); Phys.
Rev. {\bf 124}, 246 (1961).

\bibitem{gn}
 D.J. Gross and A. Neveu, Phys. Rev. D {\bf 10}, 3235 (1974).

\bibitem{Krive:1987cr}
  I.V.~Krive and A.S.~Rozhavsky,
  Sov.\ Phys.\ Usp.\  {\bf 30}, 370 (1987)
  [Usp.\ Fiz.\ Nauk {\bf 152}, 33 (1987)].

\bibitem{n2a}
A.J. Heeger, S. Kivelson, J.R. Schrieffer , and W.-P. Su, Rev. Mod.
Phys. {\bf 60}, 781 (1988).


\bibitem{Chodos:1997pg}
A. Chodos and H. Minakata, Phys. Lett. A {\bf 191}, 39 (1994);
  Lect.\ Notes Phys.\  {\bf 508}, 231 (1998)
  [hep-th/9709197];
H.~Caldas, J.L.~Kneur, M.B.~Pinto and R.O.~Ramos,
  Phys.\ Rev.\  B {\bf 77}, 205109 (2008).

\bibitem{n2c}
V. Sch\"{o}n and M. Thies, At the Frontier of Particle Physics:
Handbook of QCD: ''Boris Ioffe Festschrift'', Vol. {\bf 3}, 1945,
World Scientific (2001) [hep-th/0008175];
M.~Thies, J.\ Phys.\ A {\bf 39}, 12707 (2006).


\bibitem{Semenoff:1998bk}
  G.W.~Semenoff, I.A.~Shovkovy and L.C.R.~Wijewardhana,
  Mod.\ Phys.\ Lett.\ A {\bf 13}, 1143 (1998).

\bibitem{zkke}
V.C.~Zhukovsky, K.G.~Klimenko, V.V.~Khudyakov and D.~Ebert,
  JETP Lett.\  {\bf 73}, 121 (2001);
V.C.~Zhukovsky and K.G.~Klimenko,
  Theor.\ Math.\ Phys.\  {\bf 134}, 254 (2003);
E.J.~Ferrer, V.P.~Gusynin and V.~de la Incera,
  Mod.\ Phys.\ Lett.\ B {\bf 16}, 107 (2002);
  Eur.\ Phys.\ J.\ B {\bf 33}, 397 (2003).

\bibitem{marino}
E.C.~Marino and L.H.C.M.~Nunes, Nucl.\ Phys.\ B {\bf 741}, 404
(2006); L.H.C.M.~Nunes, R.L.S. Farias and E.C.~Marino, Phys. Lett. A
{\bf 376}, 779 (2012).

\bibitem{kzz}
 K.G.~Klimenko, R.N.~Zhokhov and V.C.~Zhukovsky,
  Phys.\ Rev.\ D {\bf 86}, 105010 (2012).

\bibitem{kzz2}
K.G.~Klimenko, R.N.~Zhokhov and V.C.~Zhukovsky, Mod.\ Phys.\ Lett.\ A {\bf 28}, 1350096 (2013).

\bibitem{n21}
D. Mesterhazy, J. Berges, and L. von Smekal,  Phys. Rev. B {\bf
  86}, 245431 (2012).

\bibitem{Ebert:2015hva}
  D.~Ebert, K.G.~Klimenko, P.B.~Kolmakov and V.C.~Zhukovsky,
  arXiv:1509.08093 [cond-mat.mes-hall].

\bibitem{n11}
K.G. Klimenko, Z. Phys. C {\bf 37}, 457 (1988);
B. Rosenstein, B.J. Warr, and S.H. Park, Phys. Rev. D {\bf 39},
3088 (1989); Phys. Rev. Lett. {\bf 62}, 1433 (1989).

\bibitem{24}
G.W.~Semenoff and L.C.R.~Wijewardhana,
 Phys. Rev. Lett.  {\bf 63}, 2633 (1989);
  Phys. Rev. D {\bf 45}, 1342 (1992).

\bibitem{28}
B.~Rosenstein, B.J.~Warr and S.H.~Park,
Phys.\ Rept.\  {\bf 205}, 59 (1991).

\bibitem{klimenko}
K.G.~Klimenko,
  Z.\ Phys.\ C {\bf 54}, 323 (1992);
  Theor.\ Math.\ Phys.\  {\bf 89}, 1161 (1992);
  Theor.\ Math.\ Phys.\  {\bf 90}, 1 (1992);
V.P.~Gusynin, V.A.~Miransky and I.A.~Shovkovy,
  Phys.\ Rev.\ Lett.\  {\bf 73}, 3499 (1994);
A.S.~Vshivtsev, B.V.~Magnitsky and K.G.~Klimenko,
  Phys.\ Atom.\ Nucl.\  {\bf 57}, 2171 (1994);
  Theor.\ Math.\ Phys.\  {\bf 106}, 319 (1996);
  V.P.~Gusynin, D.K.~Hong and I.A.~Shovkovy,
  Phys.\ Rev.\ D {\bf 57}, 5230 (1998).

\bibitem{chodos}
A.~Chodos, H.~Minakata, F.~Cooper, A.~Singh, and W.~Mao,
  Phys. Rev. D {\bf 61}, 045011 (2000).

\bibitem{toki}
D.~Ebert, K.G.~Klimenko and H.~Toki,
  Phys.\ Rev.\ D {\bf 64}, 014038 (2001);
H.~Kohyama, Phys.\ Rev.\ D {\bf 77}, 045016 (2008);
  Phys.\ Rev.\ D {\bf 78}, 014021 (2008).


\bibitem{gubina1}
D.~Ebert, K.G.~Klimenko, A.V.~Tyukov and V.C.~Zhukovsky,
  Phys.\ Rev.\ D {\bf 78}, 045008 (2008);
D.~Ebert, K.G.~Klimenko,  Phys.\ Rev.\  {\bf D80}, 125013 (2009);
D.~Ebert, N.V.~Gubina, K.G.~Klimenko, S.G.~Kurbanov and V.C.~Zhukovsky,
  Phys.\ Rev.\ D {\bf 84}, 025004 (2011).

\bibitem{gubina}
N.V.~Gubina, K.G.~Klimenko, S.G.~Kurbanov and V.C.~Zhukovsky,
  Phys.\ Rev.\ D {\bf 86}, 085011 (2012);
Moscow Univ.\ Phys.\ Bull.\  {\bf 67}, 131 (2012).

\bibitem{kneur}
J.-L.~Kneur, M.B.~Pinto, R.O.~Ramos and E.~Staudt,
  Phys.\ Rev.\ D {\bf 76}, 045020 (2007);
  Phys.\ Lett.\ B {\bf 657}, 136 (2007).

\bibitem{k}
K.G.~Klimenko,
  Z.\ Phys.\ C {\bf 50}, 477 (1991);
  Mod.\ Phys.\ Lett.\ A {\bf 9}, 1767 (1994).

\bibitem{oj}
  I.~Ojima and R.~Fukuda,
  Prog.\ Theor.\ Phys.\  {\bf 57}, 1720 (1977).

\bibitem{vas}
  A.N.~Vasiliev and G.Y.~Panasyuk,
  Theor.\ Math.\ Phys.\  {\bf 103}, 570 (1995)
  [Teor.\ Mat.\ Fiz.\  {\bf 103}, 295 (1995)].

\bibitem{thies1}
M.~Thies,
  Phys.\ Rev.\ D {\bf 68}, 047703 (2003);
 Phys.\ Rev.\ D {\bf 90}, no. 10, 105017 (2014).

\bibitem{ekkz}
D.~Ebert, T.G.~Khunjua, K.G.~Klimenko and V.C.~Zhukovsky,
  Phys.\ Rev.\ D {\bf 90}, 045021 (2014).

\bibitem{coleman}
N.D. Mermin and H. Wagner, Phys.\ Rev.\ Lett. {\bf 17}, 1133 (1966);
S. Coleman, Commun. Math. Phys. {\bf 31}, 259 (1973).

\bibitem{andrianov}
 A.A.~Andrianov, D.~Espriu and X.~Planells,
  Eur.\ Phys.\ J.\ C {\bf 73}, 2294 (2013);
 Eur.\ Phys.\ J.\ C {\bf 74}, 2776 (2014);
R.~Gatto and M.~Ruggieri,
  Phys.\ Rev.\ D {\bf 85}, 054013 (2012);
M.~Ruggieri,
  arXiv:1110.4907;
 L.~Yu, H.~Liu and M.~Huang,
 Phys.\ Rev.\ D {\bf 90}, 074009 (2014);
L.~Yu, H.~Liu and M.~Huang,
  arXiv:1511.03073 [hep-ph];
G.~Cao and P.~Zhuang,
  Phys.\ Rev.\ D {\bf 92}, 105030 (2015).

\bibitem{Braguta}
V.V.~Braguta and A.Y.~Kotov,
  arXiv:1601.04957 [hep-th].

\bibitem{ms}
  V.A.~Miransky and I.A.~Shovkovy,
  Phys.\ Rept.\  {\bf 576}, 1 (2015).

\bibitem{kharzeev}
K. Fukushima, D.E. Kharzeev, H.J. Warringa,  Phys.\ Rev.\ D {\bf 78}, 074033 (2008).

\bibitem{mizher}
A.J.~Mizher, A.~Raya and C.~Villavicencio,
  Int.\ J.\ Mod.\ Phys.\ B  {\bf 30}, 1550257 (2015).

\bibitem{mudry}
S. Ryu, C. Mudry, C.-Y. Hou, and C. Chamon,  Phys.\ Rev.\ B {\bf 80}, 205319 (2009).

\bibitem{pauli}
W. Pauli, Nuovo Cimento, {\bf 6}, 204 (1957); F.
G\"ursey, Nuovo Cimento, {\bf 7}, 411, (1957).

\bibitem{Zhukovsky:2000yd}
V.C.~Zhukovsky, K.G.~Klimenko and V.V.~Khudyakov,
 Theor.\ Math.\ Phys.\  {\bf 124}, 1132 (2000)
 [Teor.\ Mat.\ Fiz.\  {\bf 124}, 323 (2000)].

\bibitem{van} 
B.~Vanderheyden and A.D.~Jackson,
  Phys.\ Rev.\ D {\bf 61}, 076004 (2000);
V.C.~Zhukovsky, V.V.~Khudyakov, K.G.~Klimenko and D.~Ebert,
  JETP Lett.\  {\bf 74}, 523 (2001)
  [Pisma Zh.\ Eksp.\ Teor.\ Fiz.\  {\bf 74}, 595 (2001)].

\bibitem{Cao}
G.~Cao, L.~He and P.~Zhuang,
  Phys.\ Rev.\ D {\bf 90}, 056005 (2014).

\bibitem{vasiliev}
A.N. Vasiliev,``Functional methods in quantum field theory  and
statistical physics'', Leningrad Univ. Press, Leningrad, 1976.

\bibitem{Ebert:2009ty}
D.~Ebert, K.G.~Klimenko,  Phys.\ Rev.\  {\bf D80}, 125013 (2009).

\end{thebibliography}
\end{document}